\documentclass[twocolumn,iop,numberedappendix,appendixfloats]{openjournal}
\usepackage[utf8]{inputenc}
\usepackage{amsmath,amssymb,amstext}
\usepackage{graphicx}
\graphicspath{ {./images/} }
\usepackage{bm}
\usepackage{natbib}
\usepackage{subcaption}
\usepackage[colorlinks,allcolors=blue]{hyperref}
\usepackage{dsfont}
\usepackage{comment}
\usepackage{float}
\usepackage{xcolor}
\usepackage{listings}
\usepackage{fontawesome}
\definecolor{maroon}{cmyk}{0, 0.87, 0.68, 0.32}
\definecolor{halfgray}{gray}{0.55}
\definecolor{ipython_frame}{RGB}{207, 207, 207}
\definecolor{ipython_bg}{RGB}{247, 247, 247}
\definecolor{ipython_red}{RGB}{186, 33, 33}
\definecolor{ipython_green}{RGB}{0, 128, 0}
\definecolor{ipython_cyan}{RGB}{64, 128, 128}
\definecolor{ipython_purple}{RGB}{170, 34, 255}

\lstdefinelanguage{iPython}{
    morekeywords={access,and,break,class,continue,def,del,elif,else,except,exec,finally,for,from,global,if,import,in,is,lambda,not,or,pass,print,raise,return,try,while},%
    morekeywords=[2]{abs,all,any,basestring,bin,bool,bytearray,callable,chr,classmethod,cmp,compile,complex,delattr,dict,dir,divmod,enumerate,eval,execfile,file,filter,float,format,frozenset,getattr,globals,hasattr,hash,help,hex,id,input,int,isinstance,issubclass,iter,len,list,locals,long,map,max,memoryview,min,next,object,oct,open,ord,pow,property,range,raw_input,reduce,reload,repr,reversed,round,set,setattr,slice,sorted,staticmethod,str,sum,super,tuple,type,unichr,unicode,vars,xrange,zip,apply,buffer,coerce,intern, function, @model, Uniform, Normal, MvNormal, theory_planck},%
    sensitive=true,%
    morecomment=[l]\#,%
    morestring=[b]',%
    morestring=[b]",%
    morestring=[s]{'''}{'''},% used for documentation text (mulitiline strings)
    morestring=[s]{"""}{"""},% added by Philipp Matthias Hahn
    morestring=[s]{r'}{'},% `raw' strings
    morestring=[s]{r"}{"},%
    morestring=[s]{r'''}{'''},%
    morestring=[s]{r"""}{"""},%
    morestring=[s]{u'}{'},% unicode strings
    morestring=[s]{u"}{"},%
    morestring=[s]{u'''}{'''},%
    morestring=[s]{u"""}{"""},%
    literate=
    {á}{{\'a}}1 {é}{{\'e}}1 {í}{{\'i}}1 {ó}{{\'o}}1 {ú}{{\'u}}1
    {Á}{{\'A}}1 {É}{{\'E}}1 {Í}{{\'I}}1 {Ó}{{\'O}}1 {Ú}{{\'U}}1
    {à}{{\`a}}1 {è}{{\`e}}1 {ì}{{\`i}}1 {ò}{{\`o}}1 {ù}{{\`u}}1
    {À}{{\`A}}1 {È}{{\'E}}1 {Ì}{{\`I}}1 {Ò}{{\`O}}1 {Ù}{{\`U}}1
    {ä}{{\"a}}1 {ë}{{\"e}}1 {ï}{{\"i}}1 {ö}{{\"o}}1 {ü}{{\"u}}1
    {Ä}{{\"A}}1 {Ë}{{\"E}}1 {Ï}{{\"I}}1 {Ö}{{\"O}}1 {Ü}{{\"U}}1
    {â}{{\^a}}1 {ê}{{\^e}}1 {î}{{\^i}}1 {ô}{{\^o}}1 {û}{{\^u}}1
    {Â}{{\^A}}1 {Ê}{{\^E}}1 {Î}{{\^I}}1 {Ô}{{\^O}}1 {Û}{{\^U}}1
    {œ}{{\oe}}1 {Œ}{{\OE}}1 {æ}{{\ae}}1 {Æ}{{\AE}}1 {ß}{{\ss}}1
    {ç}{{\c c}}1 {Ç}{{\c C}}1 {ø}{{\o}}1 {å}{{\r a}}1 {Å}{{\r A}}1
    {€}{{\EUR}}1 {£}{{\pounds}}1
    {^}{{{\color{ipython_purple}\^{}}}}1
    {=}{{{\color{ipython_purple}=}}}1
    {+}{{{\color{ipython_purple}+}}}1
    {-}{{{\color{ipython_purple}-}}}1
    {*}{{{\color{ipython_purple}$^\ast$}}}1
    {/}{{{\color{ipython_purple}/}}}1
    {+=}{{{+=}}}1
    {-=}{{{-=}}}1
    {*=}{{{$^\ast$=}}}1
    {/=}{{{/=}}}1,
    literate=
    *{-}{{{\color{ipython_purple}-}}}1
     {?}{{{\color{ipython_purple}?}}}1,
    identifierstyle=\color{black}\ttfamily,
    commentstyle=\color{ipython_cyan}\ttfamily,
    stringstyle=\color{ipython_red}\ttfamily,
    keepspaces=true,
    showspaces=false,
    showstringspaces=false,
    rulecolor=\color{ipython_frame},
    frameround={t}{t}{t}{t},
    numbers=none,
    numberstyle=\tiny\color{halfgray},
    backgroundcolor=\color{ipython_bg},
    basicstyle=\ttfamily\footnotesize,
    columns=fullflexible,
    keywordstyle=\color{ipython_green}\ttfamily,
}

\newcommand{\blast}{\texttt{Blast.jl}}
\newcommand{\threebytwo}{$3\times2$ pt}
\newcommand{\julia}{\texttt{Julia}}
\newcommand{\nk}{\texttt{N5K}}

\newcommand{\github}{\href{https://github.com/sofiachiarenza/blast_paper}{\faGithub}}

\date{\today}

\begin{document}
\journalinfo{The Open Journal of Astrophysics}
\submitted{submitted October 2024; accepted December 2024}

\shorttitle{\blast{}: A fast and efficient algorithm for \threebytwo{} analysis.}
\shortauthors{}
\title{\blast{}: Beyond Limber Angular power Spectra Toolkit. A fast and efficient algorithm for \threebytwo{} analysis.}
\author{Sofia Chiarenza$^{\star1,2,3}$}
\author{Marco Bonici$^{1,2}$}
\author{Will J. Percival$^{1,2,3}$}
\author{Martin White$^{4,5,6}$}
\affiliation{$^1$ Waterloo Centre for Astrophysics, University of Waterloo, Waterloo, ON N2L 3G1, Canada}
\affiliation{$^2$ Department of Physics and Astronomy, University of Waterloo, Waterloo, ON N2L 3G1, Canada}
\affiliation{$^3$ Perimeter Institute for Theoretical Physics, 31 Caroline St North, Waterloo, ON N2L 2Y5, Canada}
\affiliation{$^4$ Berkeley Center for Cosmological Physics, UC Berkeley, CA 94720, USA}
\affiliation{$^5$ Department of Physics, University of California, Berkeley, CA 94720, USA}
\affiliation{$^6$ Lawrence Berkeley National Laboratory, One Cyclotron Road, Berkeley, CA 94720, USA}
\thanks{$^\star$ E-mail: \nolinkurl{schiaren@uwaterloo.ca} }

\begin{abstract}
The advent of next-generation photometric and spectroscopic surveys is approaching, bringing more data with tighter error bars. As a result, theoretical models will become more complex, incorporating additional parameters, which will increase the dimensionality of the parameter space and make posteriors more challenging to explore. Consequently, the need to improve and speed up our current analysis pipelines will grow. In this work, we focus on the \threebytwo{} statistics, a summary statistic that has become increasingly popular in recent years due to its great constraining power. These statistics involve calculating angular two-point correlation functions for the auto- and cross-correlations between galaxy clustering and weak lensing. The corresponding model is determined by integrating the product of the power spectrum and two highly-oscillating Bessel functions over three dimensions, which makes the evaluation particularly challenging. Typically, this difficulty is circumvented by employing the so-called Limber approximation, which is an important source of error. We present \blast{}, an innovative and efficient algorithm for calculating angular power spectra without employing the Limber approximation or assuming a scale-dependent growth rate, based on the use of Chebyshev polynomials. The algorithm is compared with the publicly available beyond-Limber codes, whose performances were recently tested by the Rubin Observatory Legacy Survey of Space and Time Dark Energy Science Collaboration. At similar accuracy, \blast{} is $\approx 10$-$15 \times$ faster than the winning method of the challenge, also showing excellent scaling with respect to various hyper-parameters. \blast{} is publicly available on GitHub, and we release a repository where we explain how to use the code \github. 
\end{abstract}

\maketitle
\section{Introduction}\label{sec:introduction}
Next-generation spectroscopic and photometric surveys, such as Euclid \citep{euclidcollaboration2024euclidiovervieweuclid}, Rubin Observatory Legacy Survey of Space and Time (LSST) \citep{Ivezi__2019}, the Nancy Grace Roman Space Telescope \citep{wang2022high} and Dark Energy Spectroscopic Instrument (DESI) \citep{levi2019dark}, aim to achieve high-accuracy measurements of Large Scale Structure (LSS). These surveys measure various tracers, including galaxy density and weak lensing, across wider areas of the sky and at increasingly deep redshifts. Traditional cosmological parameter estimation relies on the power spectrum to compress the data into a reduced form, containing the majority of the desired information. For surveys that rely on photometric redshifts, there is little radial information, and the angular power spectra ($C_\ell$) within and between different tomographic redshift bins are sufficient to retain relevant information. This is the main idea behind \threebytwo{} statistics, which apply this technique for different combinations of tracers: galaxy clustering, lensing shear-shear correlation, and galaxy-galaxy lensing cross-correlation. A major drawback that comes with using \threebytwo{} statistics is the numerical computation of the model, which requires the calculation of a highly oscillating three-dimensional integral. A common work-around is the Limber approximation \citep{limber1953analysis} which assumes that modes between structures at different epochs do not contribute to the line-of-sight integration. This approximation reduces the highly-oscillatory $3D$ integrand into a smooth $1D$ integrand, enabling faster computation. However, it introduces significant errors, particularly on large angular scales (low $\ell$'s). Therefore, a more accurate evaluation of the \threebytwo{} model is essential to better control systematics and to preserve the information gain from new surveys.
We need not only a precise computation of the \threebytwo{} data vector but also a method that is as fast as possible. In fact, if this model is to be used to perform parameter inference through Monte Carlo Markov Chains (MCMC), we expect to perform millions of evaluations, so the task of speeding up the computation of the full non-Limber integral is very important as it will make cosmological analyses with \threebytwo{} computationally feasible. As an example, in the recent Euclid overview paper \citep{euclidcollaboration2024euclidiovervieweuclid}, \threebytwo{} statistics were computed using the Limber approximation and the MCMC took $0.5$ million CPU hours to converge.\\

In light of this problem, the LSST Dark Energy Science Collaboration \citep{Ivezi__2019} in 2020 launched the Non-local No-Nonsense Non-Limber Numerical Knockout (\nk{}) Challenge \citep{leonard2022n5k}, with the aim of assessing what the state of the art of non-Limber integration is, and determine which is the best method to incorporate in their pipeline. 
In this work we present a novel algorithm, implemented in the \blast{}\footnote{\url{https://github.com/sofiachiarenza/Blast.jl}} code, for non-Limber integration, as a late entry to the challenge. \blast{} builds on the idea of performing a decomposition of the power spectrum on the convenient basis of the Chebyshev polynomials. Thanks to this decomposition, the hardest part of the integral becomes cosmology-independent and can therefore be pre-computed only once. \\
The paper is structured as follows: in Sec.~\ref{sec:theory} the model for \threebytwo{} statistics is worked out in detail. The Limber approximation and its regimes of validity are also discussed. In Section~\ref{sec:method} our new method, which is based on the decomposition of the power spectrum on the basis of Chebyshev polynomials, is introduced. Sec.~\ref{sec:challenge} outlines the details of the \nk{} challenge, discussing the set-up, the specific configurations, the evaluation metric, and the three algorithms that took part in the challenge. We then present how \blast{} performs and scales with respect to different parameters by rigorously comparing it to the other entries in Sec.~\ref{sec:results} and \ref{sec:Discussion}. Conclusions are drawn in Sec.~\ref{sec:conclusions}.

\section{Theory}
\label{sec:theory}
\subsection{Angular Power Spectra}
To derive the model for the \threebytwo{} analysis, we start by noting that any observable can be expanded using spherical harmonics for scalar fields, such as the matter overdensity field $\delta(\mathbf{x})$. For more general cases, such as the lensing shear—a spin-$2$ field—spin-weighted spherical harmonics are used instead. In this section, we demonstrate how the angular power spectrum coefficients are derived for scalar fields on the sphere. The same approach applies to spin-$s$ fields, with the standard spherical harmonics $Y_{\ell,m}$ replaced by the spin-weighted spherical harmonics $_sY_{\ell,m}$ \citep{zaldarriaga1997all, chisari2019core}. The spherical harmonics, which provide a complete and orthonormal basis for the total angular momentum operator, were first introduced in this context by \cite{hu1997cmb}. 
A generic scalar quantity $\Tilde{A}(\mathbf{\hat{n}})$, $\mathbf{\hat{n}}$ representing the unit vector pointing toward a given direction on the sky, is related to a $3$D quantity $A(\mathbf{r},z)$ through a projection kernel $W_i(z)$:
\begin{equation}
    \tilde{A}(\mathbf{\hat{n}}) = \int \mathrm{d}z W_i(z)A(\chi(z)\mathbf{\hat{n}}; z),
\end{equation}
$\chi(z)$ being the comoving distance. In terms of its Fourier transform, $A(\mathbf{r},z)$ can be expressed as:
\begin{equation}\label{eqn:intro1}
    A(\chi(z)\mathbf{\hat{n}};z) = \int \frac{\mathrm{d}^3k}{(2\pi)^3}A(\mathbf{k},z)e^{\mathrm{i}\mathbf{k\cdot\hat{n}}\,\chi(z)}.
\end{equation}
A plane wave can be expanded as:
\begin{equation}
    e^{\mathrm{i}\mathbf{k\cdot\hat{n}}r} = 4\pi \sum_{\ell,m} \mathrm{i}^\ell j_\ell(kr) Y^*_{\ell,m}(\mathbf{\hat{k}})Y_{\ell,m}(\mathbf{\hat{n}}),
\end{equation}
this is also known as the Rayleigh's formula. Inserting this expansion in Eq.~\eqref{eqn:intro1}, one obtains:
\begin{multline}
    A(\chi(z)\,\mathbf{\hat{n}};z) = 4\pi \sum_{\ell,m} \mathrm{i}^\ell 
    \int \frac{\mathrm{d}^3k}{(2\pi)^3} A(\mathbf{k},z) \, j_\ell(k\chi(z)) \\
    \times Y^*_{\ell,m}(\mathbf{\hat{k}}) Y_{\ell,m}(\mathbf{\hat{n}}).
\end{multline}
Putting the previous equations together, a generic quantity projected on the sky becomes:
\begin{multline}
    \tilde{A}(\mathbf{\hat{n}}) = 4\pi \sum_{\ell,m} \mathrm{i}^\ell \int \mathrm{d}z \, W_i(z) 
    \int \frac{\mathrm{d}^3k}{(2\pi)^3} A(\mathbf{k},z) \\
    \times j_\ell(k\chi(z)) Y^*_{\ell,m}(\mathbf{\hat{k}}) Y_{\ell,m}(\mathbf{\hat{n}}).
\end{multline}
So, the coefficients of the spherical harmonics decomposition of $\Tilde{A}(\mathbf{\hat{n}})$ are:
\begin{multline}
    A_{\ell,m} = 4\pi \mathrm{i}^\ell \int \mathrm{d}z \, W_i(z) 
    \int \frac{\mathrm{d}^3k}{(2\pi)^3} A(\mathbf{k},z) \, j_\ell(k\chi(z)) \\
    \times Y^*_{\ell,m}(\mathbf{\hat{k}}).
\end{multline}
Therefore, if we have two scalar fields defined on the sphere $\Tilde{A}(\mathbf{n})$ and $\Tilde{B}(\mathbf{n})$, it is possible to expand both of them in spherical harmonics and compute the angular cross-correlation power spectrum \citep{jeong2009galaxy, loverde2008extended, hu1997cmb}:
\begin{multline}
    C_{\mathrm{AB}}(\ell) = \langle A_{\ell,m} B_{\ell,m}^* \rangle = \frac{2}{\pi} \int \mathrm{d}z \, W_i(z) 
    \int \mathrm{d}z' \, W_j(z') \\
    \times \int \mathrm{d}k \, k^2 P_{AB}(k,z,z') \, j_\ell(k\chi(z)) j_\ell(k\chi(z')).
\end{multline}
Here, $j_\ell(x)$ are the spherical Bessel functions\footnote{Relativistic effects such as Redshift Space Distortions and the Integrated Sachs-Wolfe effect introduce a dependency on the derivatives of the Bessel functions $j'_\ell(x)$ and $j''_\ell(x)$.}, which are present in this expression due to the use of the plane wave expansion, $P_{\mathrm{AB}}(k, \chi_1, \chi_2)$ is the $3$D power spectrum, containing the $2$-point cosmological information, and the orthonormality condition of spherical harmonics is used. The power spectrum, as described in detail by \citet{chisari2019core}, varies depending on the probes, since the transfer functions for matter overdensities and correlated galaxy shapes differ. However, these differences are very small, and, in practice, the matter power spectrum is commonly used. Following this convention, we adopt the matter power spectrum in this paper, consistent with the approach taken in the \nk{} challenge.
$W_{i,j}^{AB}(z)$ are called window functions: they are survey-specific objects that only depend on the background cosmology and account for the fact that a real survey measures number counts and cosmic shear in different tomographic redshift bins. \\
Finally, accounting for probe-specific dependencies on $k$, $z$ and $\ell$, and changing variables from $z$ to $\chi(z)$ (the comoving distance), the angular power spectrum becomes:
\begin{multline}\label{eqn:cls-general}
    C_{ij}^{\mathrm{AB}}(\ell) = N(\ell) \int_0^\infty \mathrm{d} \chi_1 \, W_i^{\mathrm{A}}(\chi_1) 
    \int_0^\infty \mathrm{d} \chi_2 \, W_j^{\mathrm{B}}(\chi_2) \\
    \times \int_0^\infty \mathrm{d}k \, k^2 P_{\mathrm{AB}}(k,\chi_1, \chi_2) 
    \frac{j_\ell(k\chi_1) j_\ell(k\chi_2)}{(k\chi_1)^\alpha (k\chi_2)^\beta}.
\end{multline}
The exponents $\alpha$ and $\beta$ assume different values depending on the probe that is considered, in particular $0$ for clustering and $2$ for lensing. Due to the nature of the spherical Bessel functions, which are highly oscillating and slowly damped, the evaluation of the integral is computationally challenging. 

\subsection{The Limber Approximation}\label{sec:limber}
The Limber approximation \citep{limber1953analysis} is commonly employed to simplify the integral in Eq.~\eqref{eqn:cls-general} and reduce it to a $1$D integral by approximating the spherical Bessel functions around their first peak, located at $k\chi \approx \ell +1/2$. The peak gives the largest contribution to the integral. In fact, due to the highly-oscillatory nature of the functions, there are large cancellations that cause the other contributions to be very small \citep{loverde2008extended}. We can then write:
\begin{equation}
    j_\ell(x) \approx \sqrt{\frac{\pi}{2\ell+1}}\delta_\mathrm{D}\left(\ell+\frac{1}{2}-x\right),
\end{equation}
$\delta_\mathrm{D}(x)$ being the Dirac delta function. Another equivalent way to get this result is to consider orthogonality relations of the spherical Bessel functions:
\begin{equation}
    \int_0^{\infty} \mathrm{d} k \, k^2 j_\ell(k\chi_1)j_\ell(k\chi_2) = \frac{2\delta_\mathrm{D}(\chi_1-\chi_2)}{\pi\chi_1\chi_2}.
\end{equation}
Applying this approximation, Eq.~\eqref{eqn:cls-general} reduces to:
\begin{equation}
    C^{\mathrm{AB}}_{ij}(\ell) \approx \int_0^{\infty} \frac{\mathrm{d}\chi}{\chi^2}K_i^{\mathrm{A}}(\chi)K_j^{\mathrm{B}}(\chi)P_{\mathrm{AB}}\left(k_\ell, z\right),
\end{equation}
with 
\begin{equation}
    k_\ell \equiv \frac{\ell+1/2}{\chi}.
\end{equation}
This formula is only valid in spatially flat cosmologies, which we assume throughout this paper. This integral is one dimensional and does not contain oscillatory function, making it inherently easier and faster to compute. The regimes of validity of this approximation are well understood \citep{loverde2008extended}. The Limber approximation, which sets $\chi_1 = \chi_2$, works well especially for higher $\ell$'s, where the feature that the integrands have for $\chi_1=\chi_2$ gets increasingly sharp. In addition, for high multipoles $\ell$ the flat sky approximation is valid. Moreover, the approximation works best when the kernels $K_{i}^{\mathrm{A}}$ are broad in comoving distance (\textit{i.e.}, the approximation works best for the weak lensing sample rather than the clustering one) and for the auto-correlation within a redshift bin, as the kernels $K_i^{\mathrm{A}}$, $K_j^{\mathrm{A}}$ are superimposed in comoving distance. \\

\cite{loverde2008extended} developed a more refined approximation of the integral, the extended Limber approximation. The idea is to Taylor expand the spherical Bessel function around its primary peak and keep higher orders of the expansion. Performing this calculation and defining $\nu \equiv kr = \ell + 1/2$, one finds:
\begin{multline}
    C^{\mathrm{AB}}_{ij}(\ell) = N(\ell) \int \frac{\mathrm{d}\chi}{\chi^2} K_i(\chi) K_j(\chi) 
    P_{\mathrm{AB}}\left(k = \frac{\nu}{\chi}\right) \Biggl\{ 1 \\
    + \frac{1}{\nu^2} \Biggl[ \frac{\chi^2}{2} \left( 
    \frac{\Tilde{K}_i^{''}(\chi)}{\Tilde{K}_i(\chi)} + \frac{\Tilde{K}_j^{''}(\chi)}{\Tilde{K}_j(\chi)} \right) \\
    + \frac{\chi^3}{6} \left( 
    \frac{\Tilde{K}_i^{'''}(\chi)}{\Tilde{K}_i(\chi)} + \frac{\Tilde{K}_j^{'''}(\chi)}{\Tilde{K}_j(\chi)} \right) 
    \Biggr] \Biggr\}.
\end{multline}
Where we defined 
\begin{equation}
    \Tilde{K}(\chi) \equiv \frac{K(\chi)}{\sqrt{\chi}}.
\end{equation}
This expression shows the first-order correction to the standard Limber approximation, which brings the error from $\mathcal{O}(\ell^{-2})$ to $\mathcal{O}(\ell^{-4})$. While the higher-order Limber can help reduce the error and improve the speed of convergence for higher $\ell$, it does not necessarily make it accurate in the low-$\ell$ regime \citep{loverde2008extended}. 

\section{The \blast{} algorithm}\label{sec:method}
We propose a new algorithm to compute the full non-Limber angular power spectra for galaxy clustering, galaxy-galaxy lensing and cosmic shear, and then discuss the implementation of this idea.

\subsection{The Algorithm}
Let us start by explicitly expressing the three integrals, incorporating the appropriate $\ell$-dependent prefactors and the correct $k$-dependency. For galaxy clustering, we get: 
\begin{multline}\label{eqn:gg}
    C_{ij}^{\mathrm{gg}}(\ell) = \frac{2}{\pi} \int_0^\infty \mathrm{d} \chi_1 \, W_i^g(\chi_1) 
    \int_0^\infty \mathrm{d} \chi_2 \, W_j^g(\chi_2) \\
    \times \int_0^\infty \mathrm{d}k \, k^2 P_{\mathrm{AB}}(k,\chi_1, \chi_2) 
    j_\ell(k\chi_1) j_\ell(k\chi_2).
\end{multline}
The kernel has the form: 
\begin{equation}
    W_i^g(\chi)= H(z)n_i(z)b_g(z), 
\end{equation}
$H(z)$ is the expansion rate, $n_i(z)$ is the redshift distribution of the sample in the $i$-th clustering redshift bin, and $b_g(z)$ is the linear galaxy bias. 

For weak lensing: 
\begin{multline}\label{eqn:ss}
    C_{ij}^{\mathrm{ss}}(\ell) = \frac{2(\ell+2)!}{\pi(\ell-2)!} \int_0^\infty \mathrm{d} \chi_1 \, W_i^s(\chi_1) 
    \int_0^\infty \mathrm{d} \chi_2 \, W_j^s(\chi_2) \\
    \times \int_0^\infty \mathrm{d}k \, k^2 P_{\mathrm{AB}}(k,\chi_1, \chi_2) 
    \frac{j_\ell(k\chi_1)}{(k\chi_1)^2} \frac{j_\ell(k\chi_2)}{(k\chi_2)^2}.
\end{multline}
For the shear sample, the kernel has shape:
\begin{equation}
    W_i^s(\chi)= \frac{3H_0^2\Omega_m}{2a}\chi \int_z^\infty \mathrm{d}z' n_i(z') \frac{\chi(z')-\chi}{\chi(z')}.
\end{equation}
Finally, for the cross-correlation of the two samples, the angular power spectrum is:
\begin{multline}\label{eqn:gs}
    C_{ij}^{\mathrm{gs}}(\ell) = \frac{2}{\pi} \sqrt{\frac{(\ell+2)!}{(\ell-2)!}} 
    \int_0^\infty \mathrm{d} \chi_1 \, W_i^g(\chi_1) 
    \int_0^\infty \mathrm{d} \chi_2 \, W_j^s(\chi_2) \\
    \times \int_0^\infty \mathrm{d}k \, k^2 P_{\mathrm{AB}}(k,\chi_1, \chi_2) 
    \frac{j_\ell(k\chi_1)}{(k\chi_1)^2} j_\ell(k\chi_2).
\end{multline}
Our strategy to solve the integrals specified in Eq.~\eqref{eqn:gg}-\eqref{eqn:gs} will be the following: we will first tackle the inner $k$-integral containing the two Bessel functions. This is the bottleneck of the computation. Having solved that, the two outer integrals over $\chi_1$-$\chi_2$ will be solved with standard quadrature techniques. The algorithm uses a Chebyshev decomposition of the power spectrum to break the integral into terms that can be pre-computed. Once these terms are available, updating for any new power spectrum only requires recalculating the decomposition coefficients, which takes up very little computing time. Chebyshev polynomials are an advantageous choice of basis as, thanks to their properties, they have close to optimal approximation ability. While the computation of the Chebyshev coefficients requires a specific grid in $k$, this approach does not impose a specific grid for the other two integration variables. This is unlike other decomposition-based algorithms, such as FFTLog, which require a logarithmic spacing of the grid. The polynomials are defined as:
\begin{equation}
    T_n(\cos\phi) = \cos(n\phi),
\end{equation}
or, equivalently, by the recursion relation:
\begin{equation}
\begin{aligned}
    T_0(x) &= 1, \\
    T_1(x) &= x, \\
    T_{n+1}(x) &= 2x T_n(x) - T_{n-1}(x).
\end{aligned}
\end{equation}
They form a complete and orthonormal basis, widely used in approximation theory due to the many mathematical properties of the polynomials, which ensure high accuracy in the approximation. For an in-depth treatment of approximation theory using Chebyshev polynomials, see \cite{trefethen2019approximation}. Below, we outline the key definitions.
In general, a function that is (Lipschitz) continuous in $[-1,1]$ has a unique representation as a Chebyshev series:
\begin{equation}\label{eqn:a1}
    f(x) = \sum_{n=0}^{\infty} a_n T_n(x),
\end{equation}
where $T_n(x)$ are the Chebyshev polynomials, and the coefficients $a_n$ are given by:
\begin{equation}\label{eqn:cheb1}
    a_n \equiv \frac{2}{\pi} \int_{-1}^1\frac{f(x)T_n(x)}{\sqrt{1-x^2}}\mathrm{d}x.
\end{equation}
In practice, it is impossible to compute an infinite number of coefficients $a_n$ thus, $f(x)$ is approximated by truncating the series to a finite sum:
\begin{equation}\label{eqn:cheba1}
    \tilde{f}(x) = \sum_{n=0}^{N-1} a_n T_n(x).
\end{equation}
Alternatively, $f(x)$ can be approximated using interpolation at Chebyshev points, a set of points defined as:
\begin{equation}\label{eqn:chebpoints}
    t_n = \cos{\left(\frac{2n+1}{2N}\pi\right)}\,\,\,n=0,...\,,N-1.
\end{equation}
Those points are defined in $[-1,1]$ and have the property of being more dense close to the edges of the interval. The interpolating polynomial at $N$ Chebyshev points is given by:
\begin{equation}\label{eqn:a2}
    \hat{f}(x) = \sum_{n=0}^{N-1} c_n T_n(x).
\end{equation}
While the two approximations defined in Eq.~\eqref{eqn:cheba1} and \eqref{eqn:a2} yield similar results, they are not identical. One approximation is obtained by the truncation of the infinite series expansion on the basis of the Chebyshev polynomials, the other one by interpolation in Chebyshev points. As a consequence, $a_n$ and $c_n$ are different sets of coefficients. However, using the properties of Chebyshev polynomials, and the fact that:
\begin{equation}
    \forall n \in\{0, N-1\} \quad \hat{f}\left(t_n\right)=f\left(t_n\right),
\end{equation}
one can show \citep{trefethen2019approximation} that they are related:
\begin{align}\label{eqn:cheb2}
    c_n = a_n &+ (a_{n+2k} + a_{n+4k} + \dots) \nonumber \\ 
    &+ (a_{-n+2k} + a_{-n+4k} + \dots).
\end{align}
for $1 \leq n \leq k-1$. That is, the coefficients in the infinite series expansion are reassigned to their aliases of degree $<k$. This property explains why Chebyshev polynomials are widely used in approximation theory, as they offer greater accuracy than other methods for a fixed number of interpolation points. As shown in \citet{trefethen2019approximation}, Chebyshev polynomials are nearly optimal approximants, with a negligible difference in accuracy compared to interpolation at truly optimal points. Moreover, the slight improvement in accuracy does not justify the additional computational cost of determining and using the optimal interpolant.
In this specific problem, we approximate the power spectrum by interpolation in the Chebyshev points (Eq.~\eqref{eqn:a2}). This choice is driven by the fact that the coefficients $c_n$ can be evaluated through a Discrete Fourier Transform (DFT) \citep{press2007numerical, frigo1997fastest}:
\begin{equation}\label{eqn:dft}
    c_n = \sum_{j=0}^{N-1} f(x_j) \mathrm{e}^{2 \mathrm{i}\pi jn/N}.
\end{equation}
Computing an FFT is much more efficient than evaluating the integrals defining the coefficients $a_n$ (Eq.~\eqref{eqn:cheb1}), which justifies our approach. 

The power spectrum, a smooth and continuous function over the interval $[k_{\mathrm{min}}, k_{\mathrm{max}}]$ (which can be remapped in $[-1,1]$), in the basis of the Chebyshev polynomials becomes:
\begin{equation}\label{eqn:cheb_pk}
    P(k,\chi_1, \chi_2; \theta) \approx \sum_{n=0}^{n_{\mathrm{max}}} c_n(\chi_1,\chi_2;\theta)T_n(k),
\end{equation}
where the dependency on a generic set of cosmological parameters, $\theta$, has now been made explicit. The choice of the sampling points $k_j$ for the power spectrum will be motivated in the next section. This decomposition enables a clear separation of geometric and cosmological components in the integrals, simplifying the overall calculation. We can write:
\begin{equation}\label{eqn:pmd}
w^{\mathrm{AB}}_\ell(\chi_1, \chi_2; \theta) \equiv \sum_{n=0}^{n_{\mathrm{max}}} c_n(\chi_1,\chi_2; \theta) \Tilde{T}^{\mathrm{AB}}_{n;\,\ell}(\chi_1,\chi_2)\,,
\end{equation}
where we defined: 
\begin{equation}\label{eqn:t_tilde}
    \Tilde{T}^{\mathrm{AB}}_{n;\,\ell}(\chi_1, \chi_2) \equiv \int_{k_{\mathrm{min}}}^{k_{\mathrm{max}}} \mathrm{d}k \, f^{\mathrm{AB}}(k) T_n(k) j_\ell(k\chi_1) j_\ell(k\chi_2)\,,
\end{equation}
with:
\begin{equation}
    f^{\mathrm{AB}}(k) =\begin{cases}
        k^2 &\mathrm { AB} = \mathrm{gg}\,,\\
        1/k^2  &\mathrm { AB} = \mathrm{ss}\,,\\
        1  &\mathrm { AB }= \mathrm{gs}\,.
    \end{cases}
\end{equation}
Here $w_{\mathrm{AB}}(\ell)$, which is commonly called ``projected matter density'', represents the inner integral in $k$. As evident from Eq.~\eqref{eqn:pmd}, the dependence on the cosmological parameters is only present in the coefficients of the Chebyshev expansion of the power spectrum $c_n(\chi_1,\chi_2;\theta)$, while $\Tilde{T}^{\mathrm{AB}}_{n;\,\ell}(\chi_1,\chi_2)$ is cosmology-independent as it is the integral of the two Bessel functions against the Chebyshev polynomials. This is the key idea of the algorithm: the $\Tilde{T}^{\mathrm{AB}}_{n;\,\ell}(\chi_1,\chi_2)$ integrals are still challenging to compute for the presence of the Bessel functions, but they can be computed once-for-all. \\

The last ingredient for a successful computation of the integral is a change of variable: introducing $R\equiv\chi_2/\chi_1$, we can switch from the $\chi_1$-$\chi_2$ to the $\chi$-$R$ basis, which allows for a better sampling of the regions that most contribute to the integral, \textit{i.e.}, when $\chi_1 \approx \chi_2$ (or, equivalently, $R\approx 1$). For more details on this coordinate change, see Appendix~\ref{appendix:chi-R}. \\
In these new variables, the integral becomes:
\begin{multline}\label{eqn:final_cls}
    C_{ij}^{\mathrm{AB}}(\ell) = \int_{0}^{\infty} \mathrm{d}\chi \int_0^1 \mathrm{d}R \, \chi 
    \Bigl[ \mathcal{K}_i^{\mathrm{A}}(\chi)\mathcal{K}_j^{\mathrm{B}}(R\chi) \\
    + \mathcal{K}_j^{\mathrm{B}}(\chi)\mathcal{K}_i^{\mathrm{A}}(R\chi) \Bigr] 
    w_\ell^{\mathrm{AB}}(\chi, R\chi).
\end{multline}
with:
\begin{equation}
    \mathcal{K}_i^{\mathrm{A}}(\chi) = \begin{cases}
K_i^{\mathrm{A}}(\chi) &\text {for clustering}\,,\\
K_i^{\mathrm{A}}(\chi)/\chi^2 &\text{for lensing}\,.
\end{cases}
\end{equation}

\subsection{The implementation}
\label{sec:implem}
The challenge that we need to face is now the computation of $\Tilde{T}^{\mathrm{AB}}_{n;\,\ell}(k)$, the integrals of the two Bessel functions and the Chebyshev polynomials, with the appropriate $k$-factor for each case. Those integrals need to be computed with high accuracy as they constitute the central step of the algorithm. We performed this integration using the Clenshaw-Curtis quadrature rule \citep{clenshaw1960method} and $N = 2^{15}+1$ points in the interval $[k_{\mathrm{min}},k_{\mathrm{max}}]$. The value for $k_{\mathrm{min}}$ is $3.571 \times 10^{-4} \,\,\mathrm{Mpc}^{-1}$, and $k_{\mathrm{max}}$ is $15.385 \,\,\mathrm{Mpc}^{-1}$; the choice of those values will be motivated in Section~\ref{sec:challenge}. $N=2^{15}+1$ is the value that provides best balance between accuracy and performance for our purposes. In this setting, all the $\Tilde{T}^{\mathrm{AB}}_{n;\,\ell}(k)$ can be computed on a laptop in $\approx 1$-$2$ hours with $n_{\mathrm{max}}$, the number of Chebyshev polynomials in the power spectrum approximation, fixed to $120$. How the accuracy scales with respect to the parameter $N$ is discussed in Appendix \ref{appendix:N}. \\
As it is the central part of the algorithm, we highlight that, whichever the choice of $N$ and $n_{\mathrm{max}}$, this computation is only to be performed once. It is also worth mentioning that the projected matter densities $w_\ell^{\mathrm{AB}}(\chi,R\chi)$ that we computed were validated using two independent and well-established codes: \texttt{TWOFAST} \citep{grasshorn2018fast} and \texttt{QuadOsc} \citep{press2007numerical}.  \texttt{TWOFAST} is a \julia{} implementation of the FFTLog algorithm \citep{talman1978numerical, hamilton2000uncorrelated}, that will be discussed below. \texttt{QuadOsc} solves the problem of the highly oscillatory integrand by integrating between its zero-crossings and successively adding up the contributions. Our evaluation of the integral was found to be in excellent agreement with both codes, confirming the validity of our new approach.  \\

To obtain the angular power spectrum coefficients, the final step is to perform the two outer integrals in $\chi$ and $R=\chi_2/\chi_1$. The integrals do not present oscillatory functions so they can be performed with standard quadrature rules. For the integral in $\chi$, the Simpson rule \citep{press2007numerical} is sufficient as the integrand is a smooth function in $\chi$. To perform this integral, we placed $96$ evenly spaced points for $\chi \in (26,7000) \,\,\mathrm{Mpc}/h$. \newline
This is not true for the $R$ coordinate, in fact, the integral presents a sharp feature for $R\approx 1$ ($\chi_1\approx \chi_2$). It arises because the biggest contribution to the integral comes from regions at the same cosmic time, as one can see from Fig.~\ref{fig:chiR}. We define the points in $R$ in the interval $(-1,1)$ as the Chebyshev points, defined in Eq.~\eqref{eqn:chebpoints}. As anticipated, they have the property of being more densely distributed close to the edges of the interval, making them well-suited for this specific task. By definition, $R$ is positive, so to account for that we placed $97$ Chebyshev points in $[-1,1]$ and only kept the strictly positive ones: $48$ points in $(0,1]$. There is another advantage to this choice: since the projected matter density $w_\ell^{\mathrm{AB}}(\chi,R)$ is evaluated on Chebyshev points in $R$, we can perform the $R$-integral using the Clenshaw-Curtis quadrature rule \citep{clenshaw1960method}, a method that is based on Chebyshev polynomials and has better convergence properties than the second-order Simpson rule.The change of coordinate is key for a successful evaluation of the angular power spectra, as is shown in Appendix \ref{appendix:chi-R}. In fact, when working with the $\chi_1$-$\chi_2$ coordinates, $500$ points in both $\chi_1$ and $\chi_2$ were not enough to correctly pick up the feature, especially for high multipoles. 
The parameters of the grids were fine-tuned to find the best balance between speed and accuracy, according to the requirements specified in Sec.~\ref{sec:challenge}. \\

The last thing to mention concerns the matter power spectrum. The method we developed treats the linear, $P_{\mathrm{lin}}(k)$, and non-linear, $P_\delta(k)$, matter power spectrum as two separate components. In particular, the splitting we do is the following: 
\begin{equation}
    P_\delta(k, \chi_1, \chi_2) = P_{\mathrm{lin}}(k, \chi_1, \chi_2) + [P_\delta -P_{\mathrm{lin}}](k, \chi_1, \chi_2).
\end{equation}
For the linear component, we performed the Chebyshev decomposition as defined in Eq.~\eqref{eqn:cheb_pk} and used that approximation to evaluate the non-Limber angular power spectrum. The non-linear part, on the other hand, is only relevant on small scales, where the Limber approximation is sufficiently accurate. It was also shown by \cite{chisari2019unequal} that the non-linear contribution decays exponentially as $\propto \exp{\left(-(D(\chi_1)-D(\chi_2))^2\right)}$, justifying the use of the approximation to correct the angular power spectrum for the non-linear contribution. In the end, the coefficients are given by:
\begin{equation}
    C^{\mathrm{tot}}(\ell) = C_{\mathrm{lin}}(\ell) + C_{\delta}^{\mathrm{limb}}(\ell) - C_{\mathrm{lin}}^{\mathrm{limb}}(\ell).
\end{equation}
It is worth emphasizing that, although we are neglecting the unequal-time contributions to the non-linear matter power spectrum, nothing prevents us from using a more refined approach, such as the midpoint approximation~\citep{delaBella:2020rpq}.%\newpage
\section{The \nk{} challenge}\label{sec:challenge}
The community has made considerable efforts to develop accurate and efficient methods for computing the full non-Limber integral \citep{ levin1996fast, campagne2017angpow, schoneberg2018beyond,fang2020beyond,feldbrugge2023complexevaluationangularpower}, emphasizing the relevance of such task. Recently, the LSST DESC launched the ``\nk{} non-Limber integration challenge'' \citep{leonard2022n5k} with the goal of finding out what the state of the art for the computation of the non-Limber angular power spectra is. This work is relevant as it established some tests to check the performance of our algorithm and compare it with the other codes that participated to the challenge. 

\subsection{Challenge set-up}
The analysis configuration is meant to mimic the LSST year 10 data set. Here, we will provide a quick summary of the challenge details, for a more in-depth description consult the \nk{} paper \citep{leonard2022n5k}. 
The context is that of a fiducial flat $\Lambda$CDM cosmology. There are $10$ tomographic redshift bins for the clustering sample, and $5$ for the weak lensing one. The kernels were provided to the participants, alongside the linear and non-linear matter power spectra. All the possible auto- and cross-correlations between those bins will be considered. The range of $\ell$ in which the integration must be non-Limber is $\ell \in (2,200)$. \\
The benchmark calculation of the integral is a robustly-validated and stable brute-force integration that took approximately $60$ hours on $12$ Intel cores in parallel. The accuracy of the methods is quantified as the $\Delta \chi^2$ with respect to this benchmark:
\begin{equation}\label{eqn:dchi}
    \Delta \chi^2 \equiv \sum_b N_b \mathrm{Tr}(\Sigma_{ij}^{\mathrm{AB}}(\ell_b)^{-1} \Delta C_{ij}^{AB}(\ell_b))^2,
\end{equation}
where $b$ is the index identifying the bandpower considered, $\Delta C_{ij}^{AB}(\ell_b)$ is the difference between the benchmark prediction and the prediction from a given non-Limber method. $N_b$ is the effective number of modes in that specific bandpower, defined as:
\begin{equation}
    N_b = \frac{f_{\mathrm{sky}}}{2} \sum_{\ell \in b}(2 \ell +1).
\end{equation}
For $f_{\mathrm{sky}}$ we are using the sky fraction of LSST, \textit{i.e.} $0.4$. Finally, $\Sigma^{\mathrm{AB}}_{ij}(\ell_b)$ is defined as:
\begin{equation}
    \Sigma^{\mathrm{AB}}_{ij}(\ell_b)= \sqrt{\frac{2}{(2\ell_b+1)\Delta \ell_b f_{sky}}}\left[ C_{ij}^{\mathrm{AB}}(\ell_b)+N^{\mathrm{AB}}_{ij}(\ell_b)\right].
\end{equation}
$\Sigma^{\mathrm{AB}}_{ij}(\ell_b)$ is a $15\times15$ matrix containing the benchmark $C_\ell$'s in bandpower $b$ and the LSST shape-noise and shot-noise terms ($\epsilon_s^2/n_s$, and $1/n_l$ with $\epsilon_s=0.28$, $n_s=27$ source galaxies per arcmin$^2$, and $n_l = 40$ lens galaxies per arcmin$^2$). It can be shown \citep{hamimeche2008likelihood, Carron_2013} that the definition of $\Delta\chi^2$ in Eq.~\eqref{eqn:dchi} is equivalent to the standard $\Delta \chi^2 = \Delta \mathbf{x}^T \mathrm{Cov}^{-1} \Delta \mathbf{x}$, $\mathrm{Cov}$ being the Gaussian covariance of the angular power spectra. The required accuracy is $\Delta \chi^2 < 0.2$ for $\ell \in (2,200)$. \\

The scaling of each entry with respect to secondary metrics was also tested. The results will be discussed below. To make the comparison to the \nk{} challenge entries completely fair, all methods were run on the same machine, using the codes and scripts in the public \nk{} repository\footnote{\url{https://github.com/LSSTDESC/N5K}}. The cluster we employed is Narval, administered by the Digital Research Alliance of Canada. A Narval node has $249$ GB of memory, and is equipped with $2$ AMD Rome $7532$ @ $2.40$ GHz $256$M cache L3 CPUs.\\

Another algorithm, \texttt{AngPow} \citep{campagne2017angpow}, also leverages Chebyshev polynomials to compute the non-Limber integral. Its approach involves using the Clenshaw-Curtis quadrature rule in a smart and efficient way to optimize the computation. Although \texttt{AngPow} was used during the challenge to validate the benchmarks, it did not participate as an official entry because it currently only supports integration for clustering kernels. While both \texttt{AngPow} and \blast{} use Chebyshev polynomials, they differ in methodology. \texttt{AngPow} focuses on optimizing the use of a well-known quadrature rule to compute the full integral, whereas \blast{} develops the idea of decomposition on a polynomial basis.

\subsection{Challenge entries}
In order to give a general idea of the state of the art of the non-Limber integration and highlight the innovative aspects of our proposed approach, we will now briefly describe the three algorithms that took part in the challenge. For a more in-depth analysis, the reader is referred to \cite{leonard2022n5k}, \cite{fang2020beyond}, \cite{schoneberg2018beyond}, and \cite{levin1996fast}.

\subsubsection{\texttt{FKEM} (\texttt{CosmoLike})}
The \texttt{FKEM} algorithm \citep{fang2020beyond} approaches the $3D$ integrals in Eq.~\eqref{eqn:gg}-\eqref{eqn:gs} by performing first the integrals in $\chi_1-\chi_2$, each of them with a Bessel function. In order to perform the separation, the power spectrum is decomposed into a linear and separable component and a non-separable one:
\begin{equation}
\begin{split}
    P_\delta(k,z(\chi_1), z(\chi_2)) &= D(\chi_1)D(\chi_2)P_{\text{lin}}(k, z=0) \\
    &\quad + [P_\delta-P_{\text{lin}}](k, \chi_1,\chi_2),
\end{split} 
\end{equation}
where $D(\chi)$ is the linear growth factor. This approximation is accurate in a $\Lambda$CDM scenario, however it breaks down in alternative cosmological models and particularly when massive neutrinos are present. Each individual Bessel integral is computed using the FFTLog algorithm \citep{talman1978numerical, hamilton2000uncorrelated}. FFTLog is widely used, particularly in cosmology, and its core concept is to perform a complex power-law decomposition of a function $f(x)$, in this case the product $K_i^A(\chi)D(\chi)$:
\begin{equation}
    f(x) = \sum_{n=0}^N c_n x^{\nu_n}.
\end{equation}
When choosing $\nu_n = 2\pi \mathrm{i} \cdot n/N + \eta$, the power law becomes $x^{\nu} = \exp{(2 \pi \mathrm{i} \cdot n/N \cdot \log{x})} \cdot x^{\eta}$, which is a Fourier Transform in the variable $\log{x}$. The integrals of the product of Bessel functions and power laws have analytical solutions, which can be efficiently computed as it is possible to use the FFT algorithm thanks to the choice of $\nu_n$. For a more detailed description of \texttt{FKEM} and the FFTLog algorithm, see \citep{leonard2022n5k, fang2020beyond}. 
\subsubsection{\texttt{matter}}
The \texttt{matter} algorithm \citep{schoneberg2018beyond}, also makes use of the FFTLog algorithm. However, the strategy is slightly different: as a first step, the inner integral in $k$ is performed. The power spectrum is approximated using the complex power law decomposition described above, $P(k,\chi_1,\chi_2)= \sum c_n(\chi_1,\chi_2)k^{\nu_n}$. Then, the integrals of the power laws in $k$ can be performed analytically. Those integrals are cosmology-independent, so they can be pre-computed and stored. Only the outer integrals are to be performed each time to get the final result. The change of basis from $\chi_1$-$\chi_2$ to $\chi$-$R$ is performed. For more details about the algorithm and the hyper-parameters, refer to \cite{schoneberg2018beyond}. This algorithm falls in the same category as \blast{}, because the first integral being computed is the one in $k$. The differences between the algorithms are significant: \texttt{matter} requires a $\log(k)$ spacing for the algorithm to be applicable, while \blast{} has more flexibility in the grid choice. In addition, the FFTLog algorithm is known to suffer from ringing and aliasing \citep{hamilton2000uncorrelated}, while the Chebyshev polynomials are not affected by those numerical instabilities \citep{trefethen2019approximation}.

\subsubsection{\texttt{Levin}}
This method, extensively described in \cite{levin1996fast} and \cite{leonard2022n5k}, rephrases the problem of the highly oscillatory $3$D integral as the solution of a system of ordinary differential equations, which is obtained by solving a linear algebra problem. 
More specifically, the solution to the differential equation is obtained at collocation points and by assuming a suitable basis function. The method implemented in the \nk{} challenge solves the integral iteratively via bisection until convergence is reached.

\section{Results}\label{sec:results}
We will now discuss the performance of \blast{}, as compared to the entries of the \nk{} challenge \citep{leonard2022n5k} introduced in Sec.~\ref{sec:challenge}. To assess our algorithm in a fair manner, all the tests performed in the challenge have also been run for \blast{}. In addition, all of the code found in the public \nk{} repository were re-run on the same machine. In this Section we will present the fiducial results, while in Sec.~\ref{sec:Discussion} we will discuss how the speed and the accuracy of the algorithm are affected by various hyper-parameters of the code, in particular the number of tomographic bins, their width, $\Delta \chi^2$ requirements and number of cores available. 

\subsection{Fiducial results}
The fiducial evaluation metric of the challenge is the time to compute the full data vector in the LSST year 10 scenario (described in Sec.~\ref{sec:challenge}) to an accuracy of $\Delta \chi^2 \leq 0.2$ for $\ell < 200$ using 64 threads on a single node. However, since it was not possible for us to use the same machine employed for the challenge\footnote{The machine used is Cori, administered by National Energy Research Scientific Computing Center (NERSC). A Cori node is a Intel Xeon Processor E5-2698 v3. This machine is now dismissed.}, we used the same exact system settings (1 node, 64 threads) but on a Narval node\footnote{Narval's CPUs are AMD Rome 7532 @ 2.40 GHz 256M cache L3.}. It is worth nothing that, when run on Narval, a newer cluster, the runtime for all the challenge entries improved by a factor of $\approx 5-10$ compared to that reported in the original paper \citep{leonard2022n5k}, except for \texttt{Levin}, whose runtime hardly changed. 
After discussing the fiducial results with $64$ threads, we will also analyze how the performance scales as a function of the number of threads (Fig.~\ref{fig:time_vs_cores}).
The runtimes are presented in Table \ref{tab:fid_res}.
\begin{table}[h!]
    \centering
    \begin{tabular}{ccc}
         \hline
         Entry name & Threads & Runtime \\
         \hline
         \texttt{FKEM} & 64 & 0.05871 $\pm$ 0.00005 s \\
         \texttt{matter} & 64 & 0.5510 $\pm$ 0.0009 s \\
         \texttt{Levin} & 64 & 5.6 $\pm$ 0.1 s \\
         \blast{} & 64 & 0.00512 $\pm$ 0.00001 s \\
         \blast{} & 32 & 0.00360 $\pm$ 0.00009 s \\
         \hline
    \end{tabular}
    \caption{Comparison of runtimes for different entries and number of threads. The fiducial results for the \nk{} challenge are with $64$ cores, however \blast{} has its best performance with $32$.}
    \label{tab:fid_res}
\end{table}
The mean runtimes were computed by performing $10$ benchmarks of all the codes. The uncertainties are evaluated as the standard deviation between the ten runs, as was done in the \nk{} challenge. According to the challenge requirements, this time does not include the evaluation of the cosmology-independent $\Tilde{T}_{n;\,\ell}^{\mathrm{AB}}(\chi_1,\chi_2)$ (Eq.~\eqref{eqn:t_tilde}). 
The baseline runtime for \blast{} is $0.00512$s when using $64$ cores. However, we observed the best performance of $0.00360$s on $32$ cores. In the fiducial scenario, \texttt{FKEM} was the fastest non-Limber algorithm. In comparison, \blast{} is $\approx 10$ times faster when using $64$ cores and $\approx 15$ times faster when using only $32$ threads. 
As noticeable in Fig.~\ref{fig:time_vs_cores}, while all of the other codes have a runtime that decreases monotonically with the number of cores, \blast{}'s performance peaks at $32$ threads and then deteriorates for higher thread counts. Fig.~\ref{fig:time_vs_cores} also shows the timings for \blast{} when reducing from \texttt{Float64} to \texttt{Float32}: this does not have a significant impact on the $\Delta\chi^2$, which increases from $0.1054$ to $0.1064$, but it improves the runtime by a factor of $2$ ``for free'', which may be useful for some applications. 
Appendix \ref{appendix:MAER} reports some figures that show the differences with respect to the benchmark at the level of the individual power spectra, for various combinations of probes and redshift bins. 

\section{Discussion}\label{sec:Discussion}
In this Section, after taking a deeper look into the fiducial results, the performance of our code is tested with respect to variations of the fiducial setup, as in \cite{leonard2022n5k}. Specifically, we will vary the number and width of the tomographic bins, the computational resources available and the accuracy requirements. 

\begin{figure}[h!]
    \centering
    \includegraphics[width=\columnwidth]{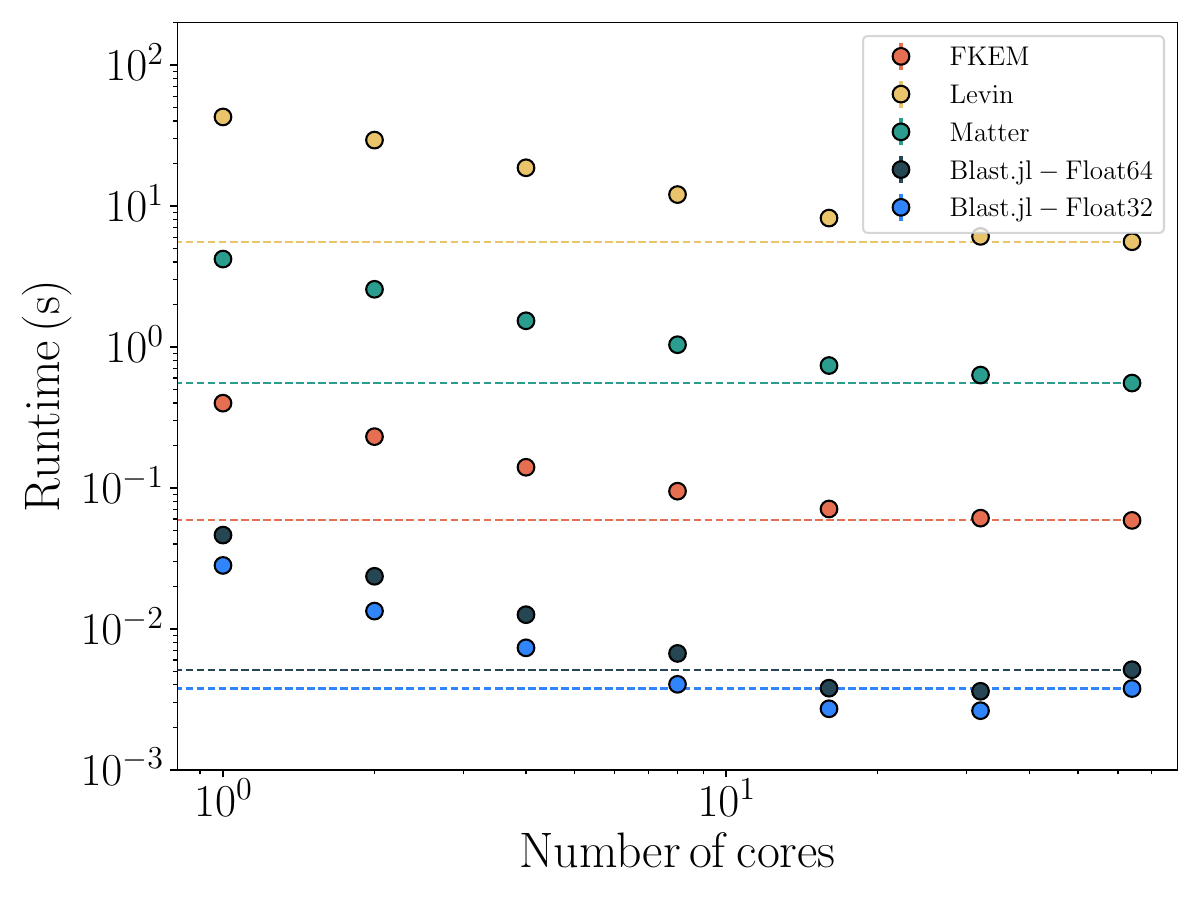}
    \caption{The impact on runtime as a function of the number of threads available on a single node. Fiducial results for 64 cores are shown as dashed horizontal lines.}
    \label{fig:time_vs_cores}
\end{figure}

Fig.~\ref{fig:time_vs_cores} shows the dependency of the runtimes on the number of computing cores employed.

\begin{figure}[h!]
    \centering
    \includegraphics[width=\columnwidth]{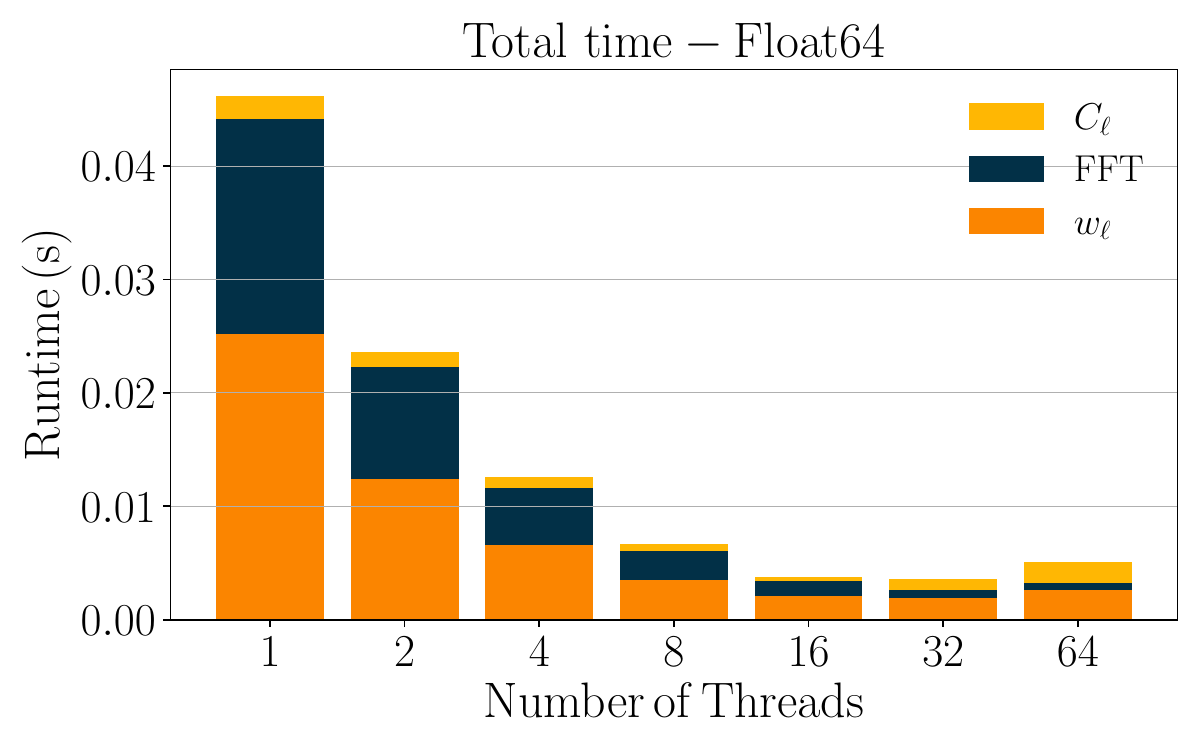}
    \caption{Breakdown of the timings for \blast{}. The bar stacked chart shows how different part of the algorithm contribute to the total time, as a function of the number of cores employed. \texttt{FFT} refers to the evaluation of the Chebyshev coefficients, $w_\ell$ is the operation in Eq.~\eqref{eqn:pmd}, and $C_\ell$ refers to the computation of the two outer integrals in $\chi$-$R$, defined in Eq.~\eqref{eqn:final_cls}.}
    \label{fig:time_breakdown}
\end{figure}

As anticipated in the previous section, \blast{} outperforms the other codes by a factor of at least $10$, which can be improved even further when working with \texttt{Float32} instead of \texttt{Float64}. This choice does not impact the $\Delta\chi^2$ significantly. A breakdown of our timings is shown in Fig.~\ref{fig:time_breakdown}, which indicates that the worse performance with $64$ threads is due to the evaluation of the projected matter densities $w_\ell$'s (defined in Eq.~\eqref{eqn:pmd}), and also to the computation of the outer integrals to get the final $C_\ell$'s. 
The reason for such a behavior is the interaction between hyper-threading and the backends used within \blast{} to perform the tensor contractions, as required to compute the $C_\ell$'s. This is not expected to be an issue, since parallelization gives almost ideal scaling up to 16 threads, and other approaches such as running more chains in parallel could be used to fully leverage the available hardware; still, we are investigating how to improve the performance when higher number of threads are employed.

\subsection{Required accuracy: runtime vs max allowed $\Delta\chi^2$}
The fiducial accuracy threshold $\Delta\chi^2<0.2$ imposed in the challenge has been chosen to ensure that any inaccuracies from non-Limber integration do not result in a false $1\sigma$ detection of a new effect described by a theory model. This conservative approach is designed to be robust even in the worst-case scenario, where the influence of non-Limber integration on the signal closely resembles that of the effect being mistakenly detected. In practice, it is likely that slightly less stringent accuracy requirements may be sufficient \citep{leonard2022n5k}. This motivated the testing of how the runtimes of the various methods scale with accuracy requirements within the range of potentially acceptable accuracy levels.
\begin{figure}[h!]%
    \centering%
    \includegraphics[width=\columnwidth]{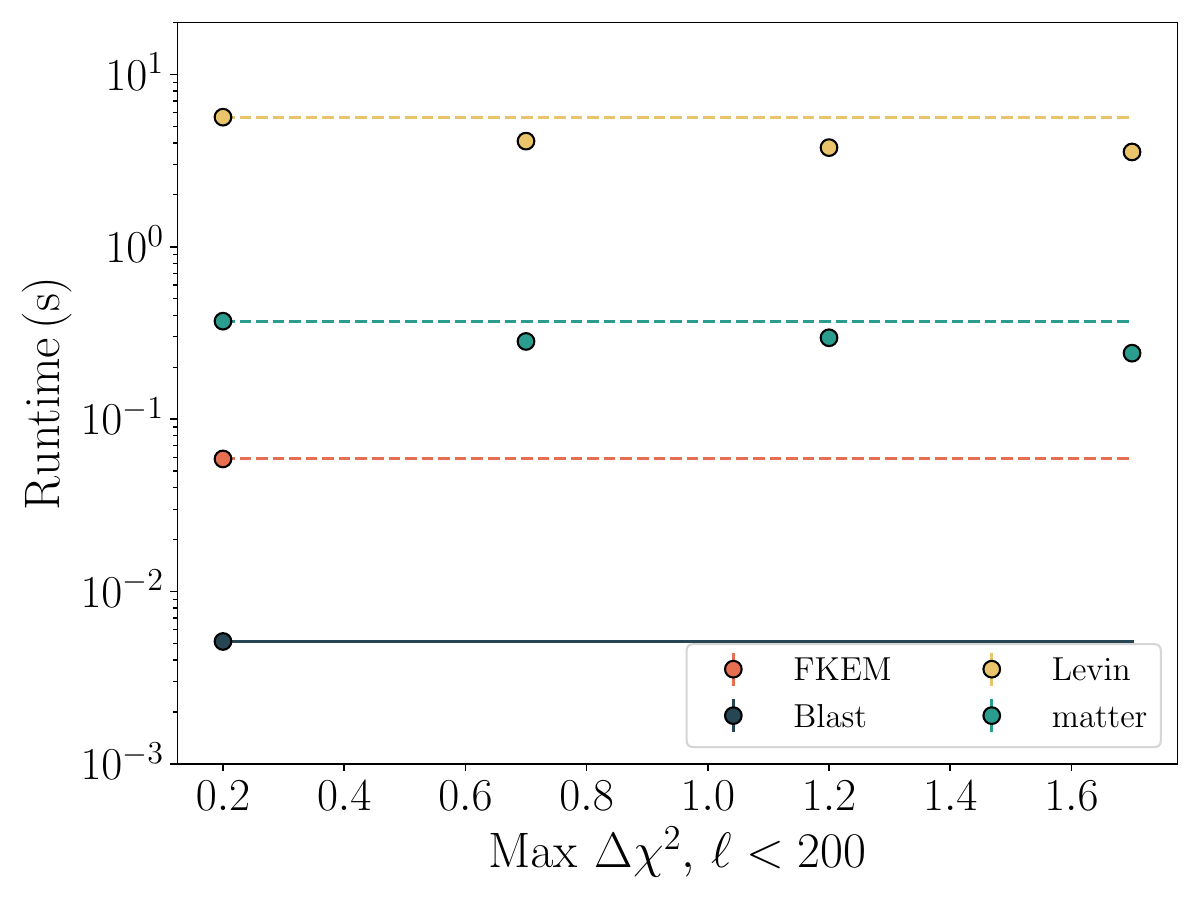}
    \caption{Run-time on 64 cores as a function of maximum allowed $\Delta\chi^2$ for $\ell <200$. Both \blast{} and \texttt{FKEM} do not present significant runtime differences on this range of allowed $\Delta\chi^2$, so we present only the fiducial $\Delta\chi^2$ point and a horizontal line for visual comparison with other methods. The fiducial case for the other two cases is also displayed as dashed lines. Uncertainties (on the mean runtimes) are too small to be visible compared to the size of points.}
    \label{fig:time_vs_dchi2}
\end{figure}
As shown in Fig.~\ref{fig:time_vs_dchi2}, four different $\Delta\chi^2$ values were tested: $\{0.2, 0.7, 1.2, 1.7\}$. To change \blast{}'s accuracy, we tweaked the number of Chebyshev polynomials used for the power spectrum approximation. If for the fiducial case ($\Delta\chi^2<0.2$) $120$ polynomials are necessary, we can reduce them to $\lesssim 100$ if we require $\Delta\chi^2<1.7$. Decreasing the number of Chebyshev polynomials by such amount does not have a significant impact on the runtime, which explains why the plot only shows the fiducial case (similarly to what happens for \texttt{FKEM}).

\subsection{Number of Spectra: Runtime vs number of tomographic bins}
\begin{figure}[h!]
    \centering
    \includegraphics[width=\columnwidth]{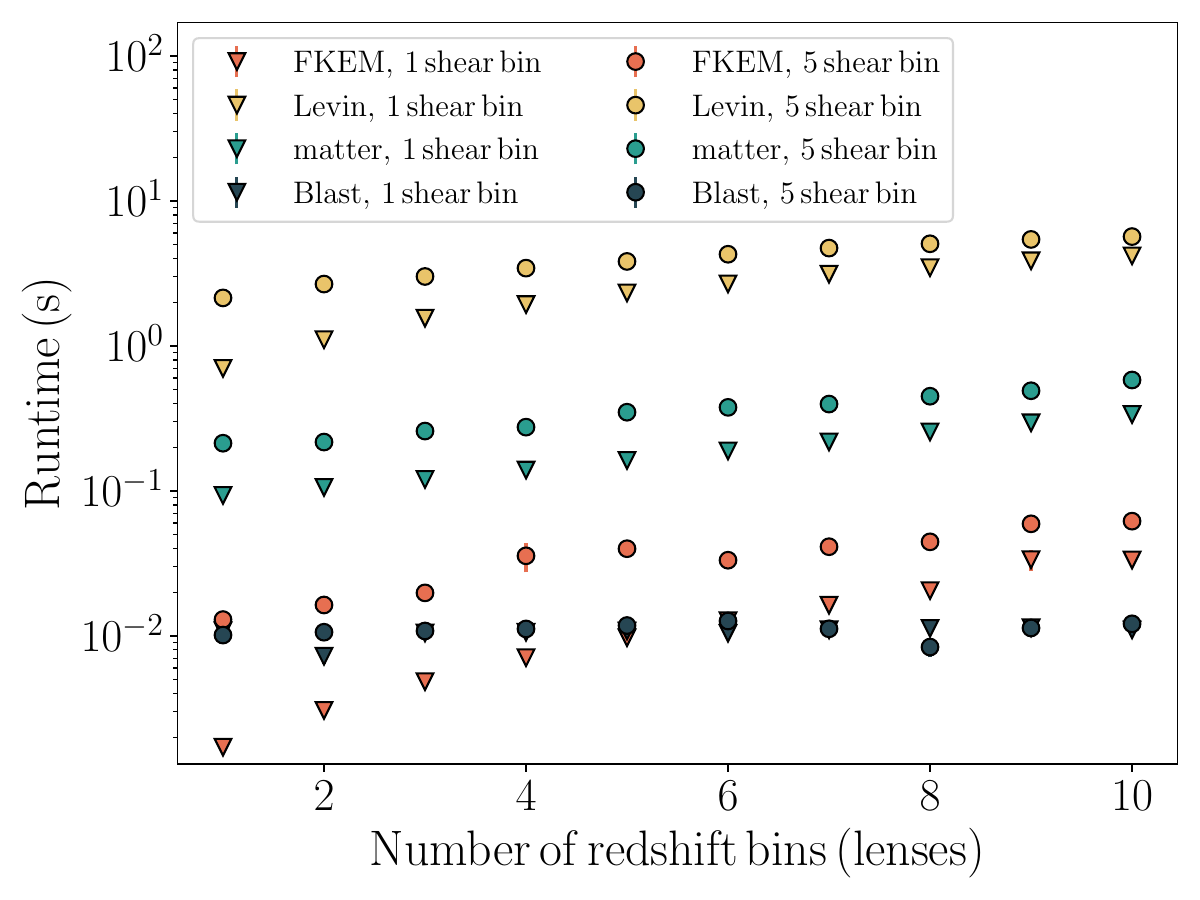}
    \caption{Timings as a function of the number of lens bins.}
    \label{fig:time_vs_nbins}
\end{figure}

As in the \nk{} challenge, we went on to examine the scaling of the runtime with the number of auto- and cross-power spectra to be computed. 
To test this, the number of shear bins was fixed to $1$ (triangular markers) and $5$ (circular markers), while the number of clustering bins was varied between $1$ and $10$. Results are shown in Fig.~\ref{fig:time_vs_nbins}. For this test, we performed $10$ benchmarks for each entry. \texttt{FKEM}, \texttt{matter} and \texttt{Levin} present a noticeable scaling with the number of tomographic bins. On the other hand, \blast{}'s results are independent of the number of bins employed in the analysis. This behavior is consistent with the results from the previous section, where the number of Chebyshev polynomials in the power spectrum approximation, varied between approximately $100$ and $120$, did not have a significant impact on the runtime. Currently, the total number of bins ranges from $2$ to $15$, which, given the structure of the code, is not expected to significantly affect the computation times.

\subsection{Width of redshift bins: $\Delta \chi^2$ vs bin width}
\begin{figure}[h!]%
    \centering%
    \includegraphics[width=\columnwidth]{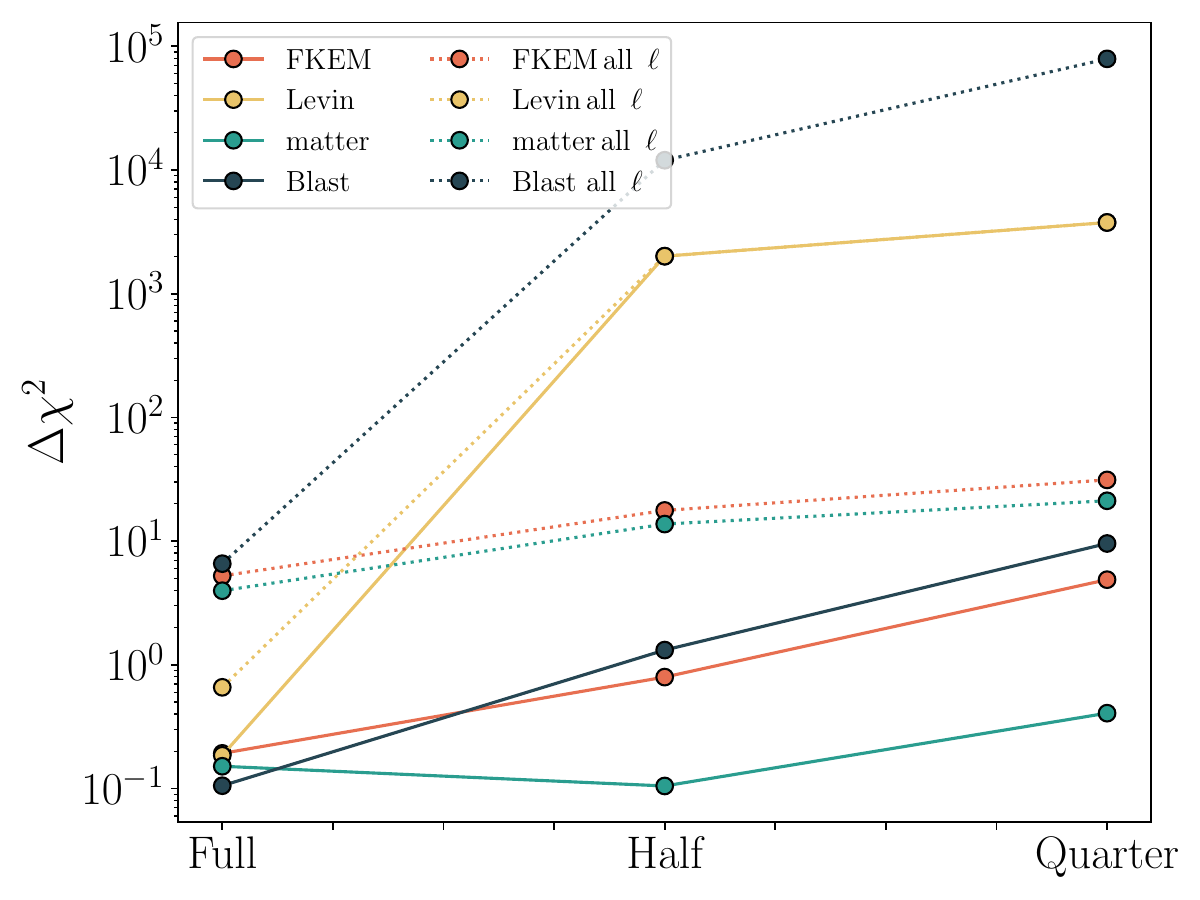}
    \caption{The impact on the accuracy for $\ell < 200$ and $\ell <2000$) achieved by each integration method as a function of the width of the clustering sample bins relative to the fiducial case.}
    \label{fig:dchi2_width}
\end{figure}

The width of the tomographic redshift bins plays a crucial role in the \threebytwo{} analysis, as it directly affects the validity of the Limber approximation (see Sec.~\ref{sec:limber}). Specifically, narrower bins lead to a greater breakdown of the Limber approximation, making accurate non-Limber computations increasingly important. Therefore, it is interesting to assess how our algorithm depends on that parameter. As in the \nk{} challenge, two new analysis scenarios were considered: in one scenario the width was reduced by a factor of $2$, in the other by a factor of $4$. Details of the new bins can be found in \cite{leonard2022n5k}. Fig.~\ref{fig:dchi2_width} shows the result of this test: the values of $\Delta\chi^2$ in the three cases (full, half and quarter width) are shown for all the codes. The continuous lines connect points for which values of $\ell <200$ are considered in the evaluation of $\Delta\chi^2$, while the dotted lines indicate that the whole $\ell$-range is considered. As expected, the reduction of the bin size causes a lower accuracy of all methods, \blast{} included. When using the fiducial settings, our algorithm performed in a very similar way as \texttt{Levin}: we observed a consistent decrease of accuracy. To improve the performance, we therefore doubled the number of sampling points in $\chi$, going from $96$ to $200$, and also increased the number of points in $R$ from $48$ to $56$. With those modifications, which match the ethos of the challenge as other entries adjusted their hyper-parameters for this test, we were able to significantly increase the accuracy, which is of the same order of magnitude as \texttt{FKEM} and \texttt{matter}. This suggests that we can push the accuracy by adding more sampling points. \\ 
The $\Delta\chi^2$ for the full $\ell$-range is much higher for \blast{} than all the other codes: this is because, at this stage, we are using the first-order Limber approximation in the range $\ell \in (200, 2000)$, while all of the other entries implement the more accurate second-order Limber, at least for $\ell \in (200, 1000)$. Since the focus of the challenge is to compute the full integral for $\ell < 200$, we did not prioritize this regime, particularly because it does not require the implementation of any new algorithm. Upon the public release of the package, higher-order Limber will be implemented for higher multipoles. 

\section{Conclusions}
\label{sec:conclusions}
The development of accurate and fast algorithms for computing cosmological quantities is of crucial importance considering the unprecedented amount and quality of data that next-generation cosmological surveys are about to bring into the field. 
In this paper we focused on improving the evaluation of the model for \threebytwo{} statistics, without the employment of the Limber approximation. We presented \blast{}, a new algorithm to compute non-Limber angular power spectra for galaxy clustering and cosmic shear samples, whose core idea is a decomposition of the power spectrum into the complete and orthonormal basis of the Chebyshev polynomials. By expressing the $3D$ matter power spectrum in terms of this basis, the most computationally challenging part of the integral can be effectively isolated, allowing for it to be pre-computed and stored. This significantly reduces the complexity of the overall computation, making it more efficient.  
The performance of our new algorithms was assessed in a fair and robust way, according to the standards of the \nk{} challenge, a non-Limber integration challenge launched in 2020 by LSST DESC \citep{Ivezi__2019} that allowed us to test \blast{} against the state-of-the-art non-Limber algorithms. 
In terms of speed of computation at the required accuracy, \blast{} outperformed \texttt{FKEM}, the winner of the challenge. In particular, our algorithm is $10$-$15\times$ faster in the fiducial scenario. \blast{} runtime also proved to be basically independent of the number of tomographic redshift bins in the analysis and is also insensitive to the accuracy requirements, making \blast{} the preferred method in those scenarios. Another scaling we explored is the accuracy as a function of the width of the redshift bins: while the performance is still comparable to that of the other challenge entries, \texttt{matter} is still the least sensitive algorithm with respect to this specific hyper-parameter, as it keeps essentially the same accuracy when the bin width is halved. Working on improving the performance of our algorithm in this scenario is reserved for future work. Nonetheless, the tests performed in this work show that increasing the number of integration points strongly impacts \blast{}'s accuracy, giving a possible solution to this behaviour. 
We also highlight that, unlike \texttt{FKEM}, \blast{} does not assume a scale-independent growth factor, so it is possible to employ it in cosmologies with massive neutrinos or modified gravity scenarios. In that sense, the algorithm we really want to compare to is \texttt{matter}. With respect to that entry, the speed in the fiducial scenario improves by over 2 orders of magnitude.  To the best of our knowledge, the fastest non-Limber integration algorithm is the one presented in this paper; actually, only the emulation of the whole \threebytwo{} statistic would result in a shorter runtime~\citep{Zhong:2024xuk, Saraivanov:2024soy}. \\

Future work will focus on adding other effects to the $C_\ell$'s, in particular redshift space distortions, magnification bias and relativistic corrections like the integrated Sachs-Wolfe effect. Those will enable the use of \blast{} for a real cosmological analysis. We will also work on including the CMB as an alternative observable, allowing the user to perform galaxy-CMB lensing cross-correlation analyses. Another future development is related to the computation of the matter power spectrum; given the fast runtime of \blast{}, it is important that this step does not become the new bottleneck of the \threebytwo{} computation. Currently, we are working on a matter power spectrum emulator, tailored to work in synergy with \blast{}.
Finally, \blast{} will also include the support for automatic differentiation. Having a \threebytwo{} statistics code that is differentiable is crucial because it enables the use of gradient-based methods \citep{Campagne_2023, Nygaard_2023, Piras_2024, Bonici_2024, Ruiz_Zapatero_2024, Balkenhol_2024, bonici2022fastemulationtwopointangular, giovanetti2024linx}. The ultimate goal is to perform cosmological parameter inference using \threebytwo{} statistic and a fully-differentiable likelihood that can be sampled very efficiently thanks to \blast{}. 

\section*{Acknowledgments}
The authors are grateful to Danielle Leonard, Elisabeth Krause, Robert Reischke, Nils Schoeneberg, David Alonso, and Jean Eric Campagne for reading the paper and providing precious feedback, and to Stefano Camera, Giulio Fabbian, Francois Lanusse, and Nicolas Tessore for useful and valuable discussions. \newline
The authors acknowledge the support of the Canadian Space Agency. WP also acknowledges support from the Natural Sciences and Engineering Research Council of Canada (NSERC), [funding reference number RGPIN-2019-03908].  MW is supported by the DOE. \newline
Research at Perimeter Institute is supported in part by the Government of Canada through the Department of Innovation, Science and Economic Development Canada and by the Province of Ontario through the Ministry of Colleges and Universities. \newline
This research was enabled in part by support provided by Compute Ontario (computeontario.ca) and the Digital Research Alliance of Canada (alliancecan.ca). 
\bibliographystyle{mnras}
\bibliography{biblio}

\clearpage
\appendix
\section{Choosing an effective basis for the integration}
\label{appendix:chi-R}
\begin{figure*}[h!]
    \centering%
    \includegraphics[width=0.8\textwidth]{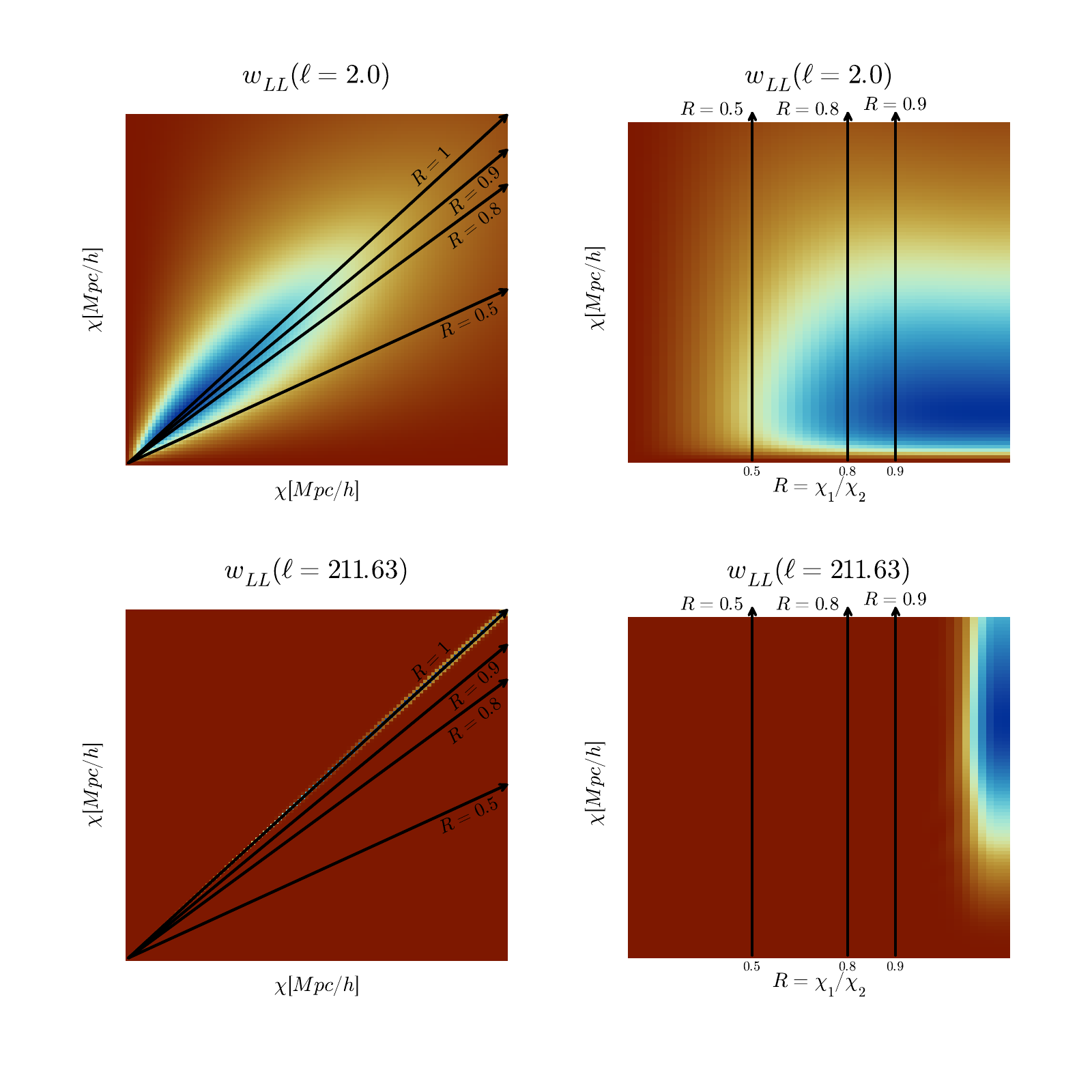}
    \caption{Visual representation of the change of the re-parametrization of the projected matter density (Eq. \eqref{eqn:pmd}) from coordinates $\chi_1$-$\chi_2$ to $\chi$-$R$. The heatmaps represent $w^{ss}(\ell = 2.0)$ and $w^{ss}(\ell = 211.63)$ in the two different coordinate systems. The image shows how defining the coordinate R allows for a better description of the object, especially for higher multipole orders. The matrices are normalized to their maximum value, so the color scheme goes from $-1$ (dark red) to $1$ (blue).}
    \label{fig:chiR}
\end{figure*}

A crucial step for the successful calculation of the angular power spectra with this method is the re-parametrization of the integral from the $\chi_1$-$\chi_2$ variables to $\chi$-$R$, with $R\equiv \chi_2 / \chi_1$. The projected matter densities (Eq. \eqref{eqn:pmd}) present a sharp feature along their diagonal ($\chi_1 = \chi_2$). Capturing this feature in the $\chi_1$-$\chi_2$ reference frame would require a high number of sampling points ($\gg500$), which is computationally not convenient. If we define $R$ as the ratio of the coordinates, we know that $R \in (0,1]$, and $R=1$ represents the diagonal. It is therefore easier and more intuitive to perform a finer sampling of the diagonal feature, and a coarser one outside the diagonal, where the contribution to the integral are close to zero. Fig. \ref{fig:chiR} is a visual representation of this change of basis, showing how most of the contributions to the integral come from $R>0.8$. Note also that the figure shows a low multipole $\ell=2$, for which the off-diagonal contributions are still relevant. For higher multipole orders, the feature gets sharper and sharper. This effectively shows why the Limber approximation, which accounts only for the contribution coming from the diagonal $\chi_1=\chi_2$, is more accurate at higher multipoles $\ell$.
The full calculation for the change of variables will now be worked out. The first step consists in noting that the integral can be symmetrized:
\begin{equation}\label{eqn:app1}
    \begin{aligned}
        C_{ij}^{\mathrm{AB}}(\ell) &= \int_0^{\infty} \mathrm{d}\chi_1 K_i^{\mathrm{A}}(\chi_1)\int_{0}^{\infty} \mathrm{d}\chi_2 K_j^{\mathrm{B}}(\chi_2)w_{\mathrm{AB}}(\chi_1,\chi_2) \\
        &= \int_0^{\infty} \mathrm{d}\chi_1 K_i^{\mathrm{A}}(\chi_1) \int_0^{\chi_1} \mathrm{d}\chi_2 K_j^{\mathrm{B}}(\chi_2)w_{\mathrm{AB}}(\chi_1,\chi_2) +\int_0^{\infty} \mathrm{d}\chi_2 K_j^{\mathrm{B}}(\chi_2) \int_0^{\chi_2} \mathrm{d}\chi_1 K_i^{\mathrm{A}}(\chi_1)w_{\mathrm{AB}}(\chi_2,\chi_1).
    \end{aligned}
\end{equation}
In the first integral, the change of variable is $R = \chi_2/\chi_1$, in the second one $R=\chi_1/\chi_2$. Adding that into Eq.~\eqref{eqn:app1}, one finds:
\begin{equation}
    C_{ij}^{\mathrm{AB}}(\ell) = \int_{0}^{\infty} \mathrm{d}\chi \int_0^1 \mathrm{d}R \, \chi \left[\mathcal{K}_i^{\mathrm{A}}(\chi)\mathcal{K}_j^{\mathrm{B}}(R\chi) + \mathcal{K}_j^{\mathrm{B}}(\chi)\mathcal{K}_i^{\mathrm{A}}(R\chi)\right]w_{\mathrm{AB}}(\chi, R\chi) = \int_{0}^{\infty} \mathrm{d}\chi \int_0^1 \mathrm{d}R \, \chi \Tilde{\mathcal{K}}_{ij}^{\mathrm{AB}}(\chi, R\chi)w_{\mathrm{AB}}(\chi, R\chi).
\end{equation}
where: 
\begin{equation}
    \mathcal{K}_i^{\mathrm{A}}(\chi) = \begin{cases}
K_i^{\mathrm{A}}(\chi) &\text{for clustering}\\
K_i^{\mathrm{A}}(\chi)/\chi^2 &\text{for lensing.}
\end{cases}
\end{equation}
\clearpage
\section{Assessing model errors}
\label{appendix:MAER}
\begin{figure*}[h!]
    \centering%
    \includegraphics[width=0.8\textwidth]{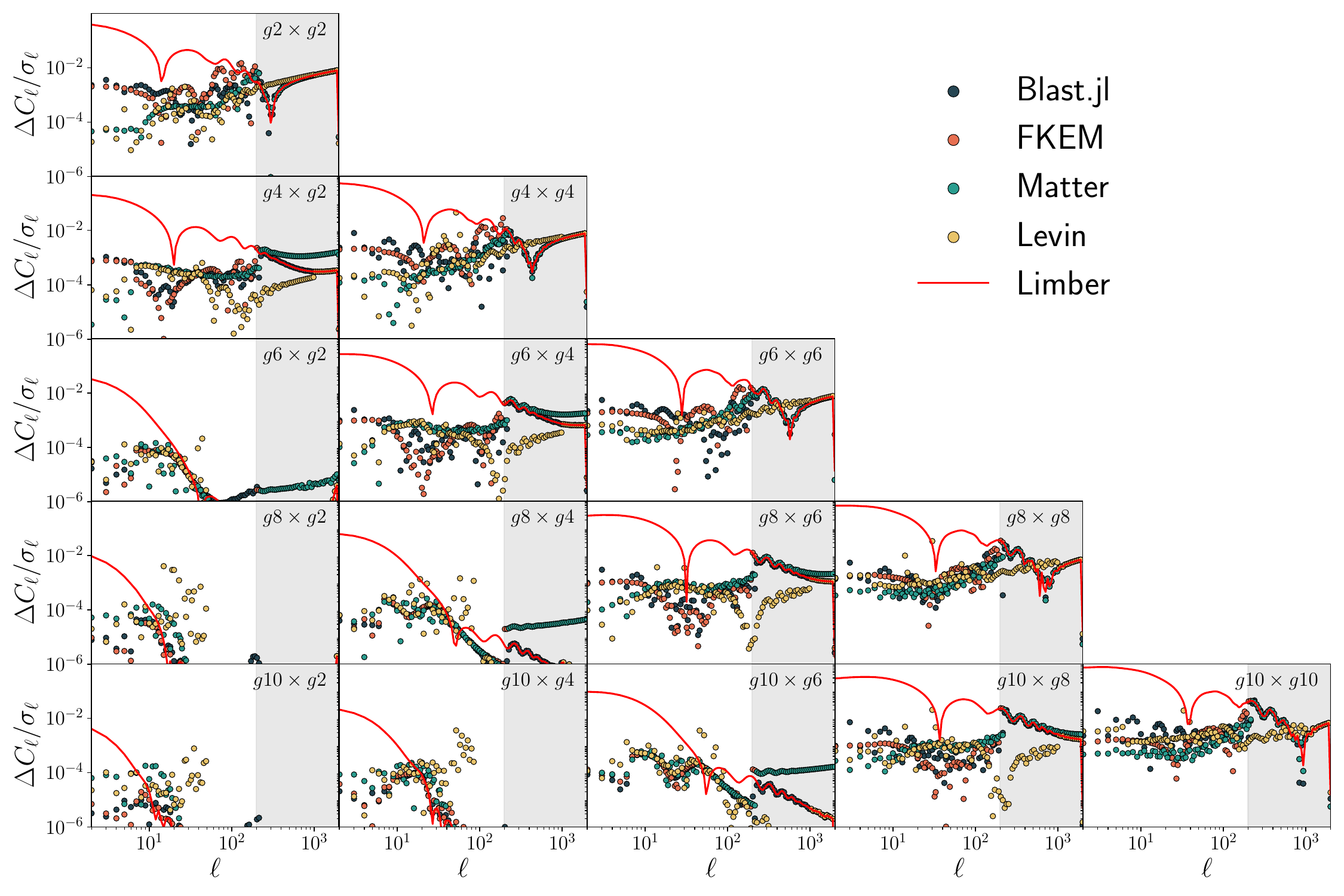}
    \caption{Deviation with respect to the benchmarks in the different auto- and cross-correlations between five of the clustering redshift bins (identified as $g$) used in this analysis as a fraction of the Gaussian uncertainties. Results are shown for the pure Limber approximation (red), Blast (blue), FKEM (orange), Levin (yellow), and matter (turquoise). The grey band represents the region where $\ell > 200$, not part of the challenge set-up. This figure visually shows how the goal of improving with respect to the Limber approximation is accomplished. Moreover, the error of our algorithm are comparable to the other challenge entries.}
    \label{fig:CC_F64}
\end{figure*}

\begin{figure*}[h!]
    \centering%
    \includegraphics[width=0.8\textwidth]{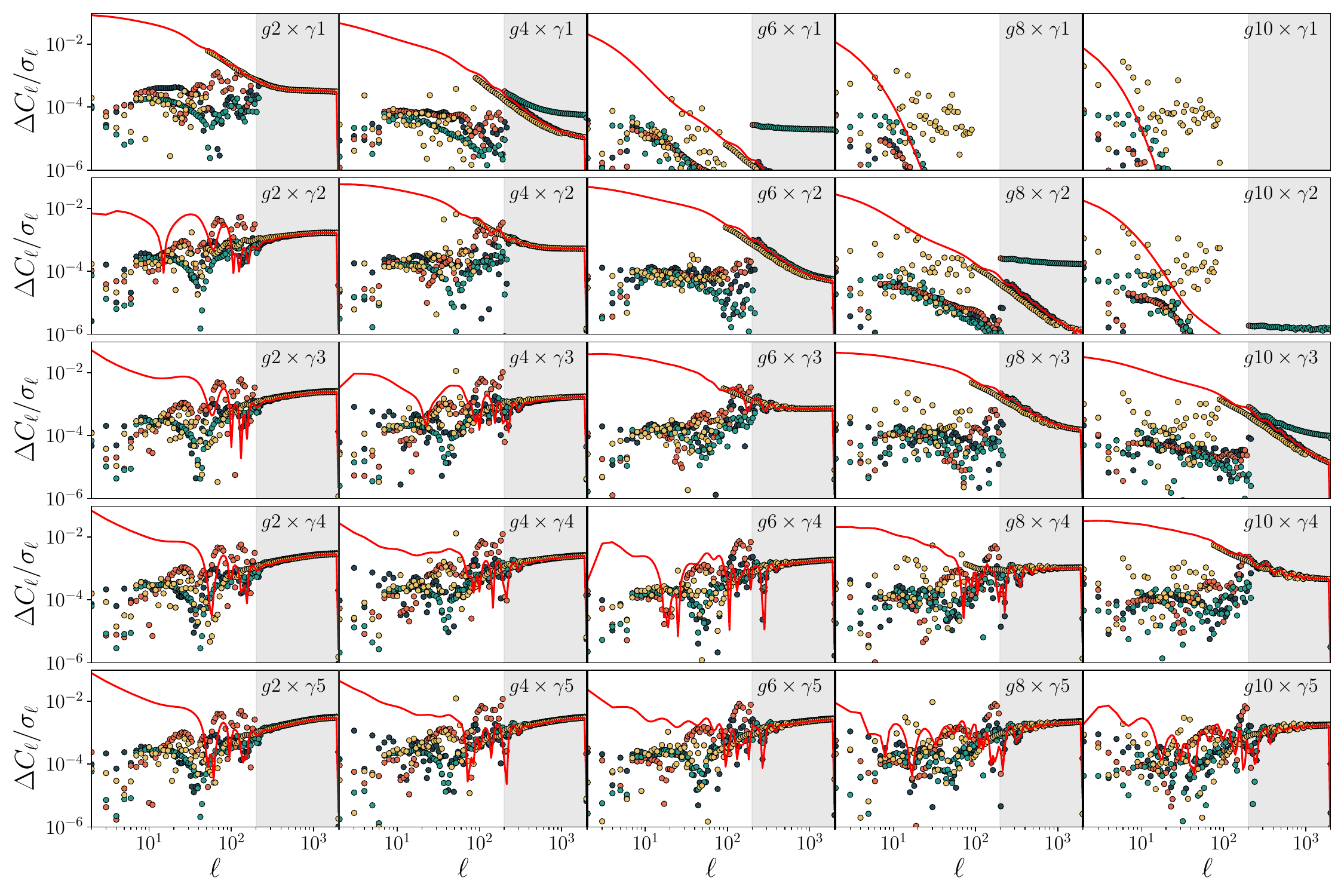}
    \caption{Same as Figure \ref{fig:CC_F64} for the clustering-shear cross-correlations (in the same clustering bins). The shear bins are identified as $\gamma$, while $g$ is used for the galaxy clustering tomographic bins.}
    \label{fig:CL_F64}
\end{figure*}

\begin{figure*}[h!]
    \centering
    \includegraphics[width=0.8\textwidth]{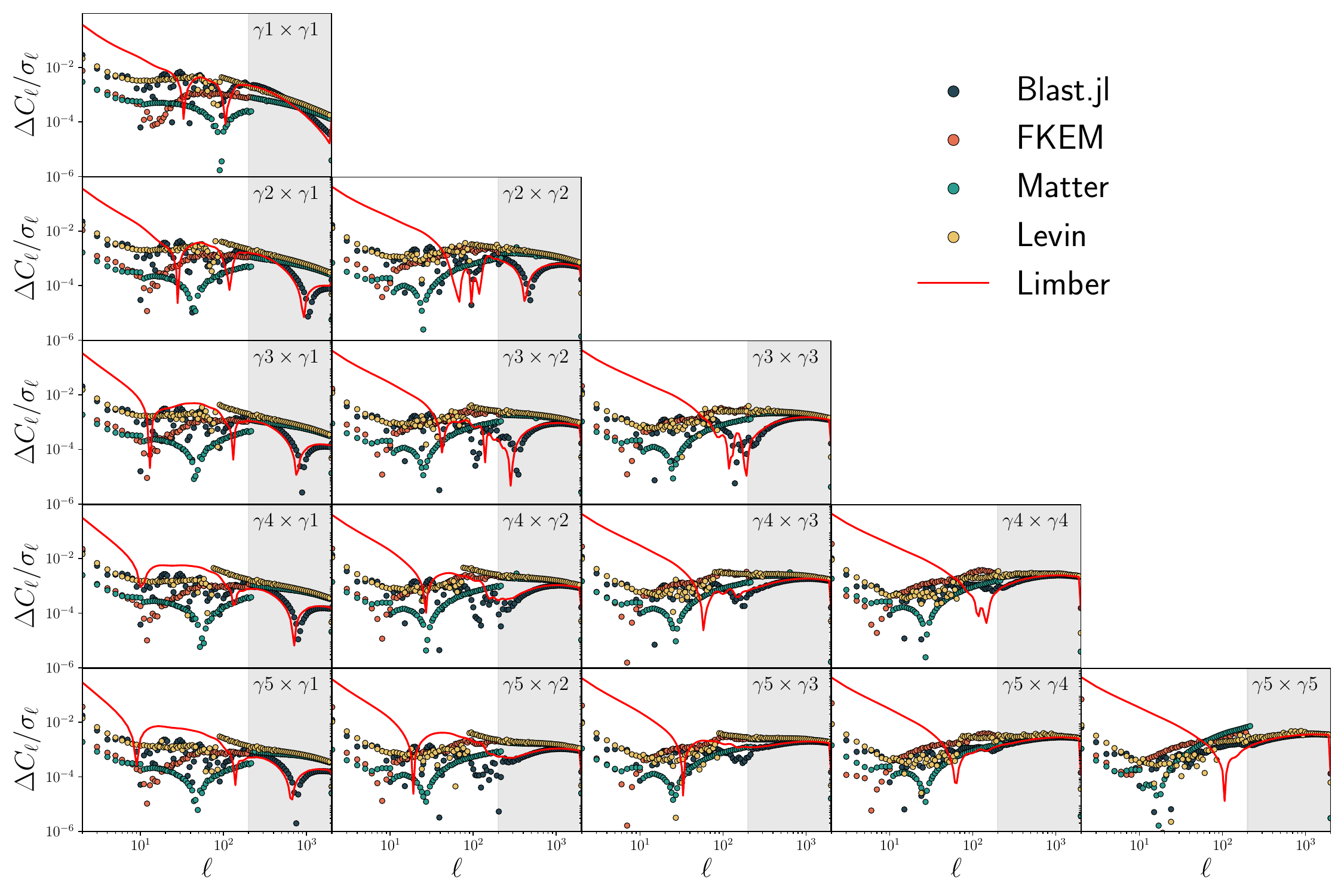}
    \caption{Same as Figure \ref{fig:CC_F64} for the shear-shear auto- and cross-correlations.}
    \label{fig:LL_F64}
\end{figure*}

The following plots show the errors with respect to the benchmarks, expressed as a fraction of the cosmic variance (the Gaussian statistical uncertainties), defined by the Knox's formula \citep{knox1995determination}:
\begin{equation}
    \sigma_{ij}^{\mathrm{AB}}(\ell) = \sqrt{\frac{C_{ii}^{\mathrm{AA}}(\ell)C_{jj}^{\mathrm{BB}}(\ell)+(C_{ij}^{\mathrm{AB}}(\ell))^2}{f_{\mathrm{sky}}(2\ell+1)}}
\end{equation}
where we are using $f_{\mathrm{sky}}=0.4$, the sky fraction for LSST. In particular, we can define a metric that is often referred to as Mean Absolute Error Ratio (MAER):
\begin{equation}\label{eqn:maer}
    \mathrm{MAER}^{\mathrm{AB}}_{ij}(\ell) = \frac{|C_{ij}^{\mathrm{AB}}(\ell)-C_{ij, \mathrm{bench}}^{\mathrm{AB}}(\ell)|}{\sigma_{ij}^{\mathrm{AB}}(\ell)}
\end{equation}
Fig.~\ref{fig:CC_F64}-\ref{fig:LL_F64} show the quantity $\Delta C_\ell / \sigma_\ell$, defined as the MAER in Eq.~\eqref{eqn:maer}, for the auto- and cross- correlations between $5$ of the clustering and shear redshift bins for the Limber approximation, the three entries of the challenge and \blast{}. As discussed in Sec.~\ref{sec:limber}, the Limber approximation is not accurate for small $\ell$, where the error is at a level of $\approx 10\%$. All the non-Limber methods, \blast{} included, are able to reduce the error on those scales. In particular, the error given by our implementation is compatible to that of the other codes, and is very stable across the whole range of $\ell$ and the different bin combinations. This is very positive as spikes in $\Delta C_\ell/\sigma$ can have a strong impact on the $\Delta\chi^2$. For $\ell > 200$, the errors are comparable to the Limber: most implementations, ours included, actually use the approximation for higher multipole orders. Finally, we discussed how the Limber approximation is more accurate for the shear-shear auto-correlation because the weak lensing kernels cover a wide range of $\chi$ values: this explains why the improvement given by all non-Limber methods is way less drastic in that case (Fig.~\ref{fig:LL_F64}).
\clearpage
\section{Pre-computation efficiency}
\label{appendix:N}
The core idea of the algorithm is to pre-compute the cosmology-independent quantities defined in Eq.~\eqref{eqn:pmd}. As discussed in Sec.~\ref{sec:implem}, we performed the integrals using the Clenshaw-Curtis integration rule and $N = 2^{15}+1$ points in the interval $[k_{\mathrm{min}}, k_{\mathrm{max}}]$. The following plot motivates the choice of the parameter $N$ by showing its impact on the accuracy, \textit{i.e.}, the $\Delta\chi^2$.
\begin{figure*}[h!]
    \centering
    \includegraphics[width=0.6\textwidth]{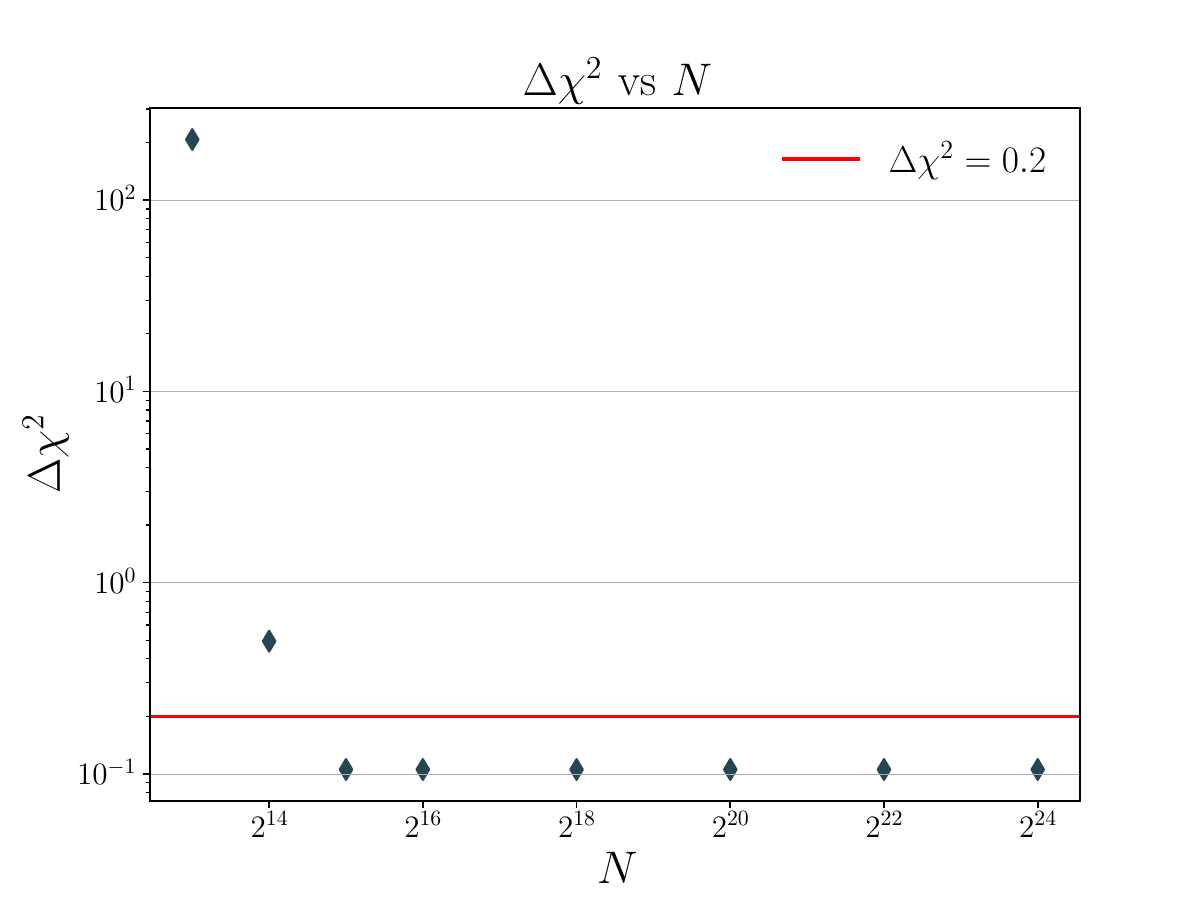}
    \caption{Accuracy as a function of the number of integration points used to evaluate the integrals in Eq.~\eqref{eqn:pmd}. The plot shows that our algorithm is well converged for $N>2^{15}$. The red line indicates the accuracy requirement from the \nk{} challenge. }
    \label{fig:app3}
\end{figure*}

\end{document}